\documentclass[a4paper,12pt]{article}
\pdfoutput=1
\usepackage[english]{babel}
\usepackage{epsfig}
\usepackage{amssymb,subfigure}
\usepackage{amsfonts}
\usepackage{amsmath}
\usepackage{euscript}
\usepackage{verbatim}
\usepackage{latexsym}
\usepackage{graphicx,color}
\usepackage{caption}
\usepackage{float}
\usepackage{ytableau}
\usepackage{hyperref}
\usepackage{cite}
\definecolor{link}{rgb}{.8,.15,.1}
\hypersetup{colorlinks=true,linkcolor=link,citecolor=link,urlcolor=link,linktocpage}
\usepackage{wrapfig}
	\usepackage{tikz}
	\usetikzlibrary{decorations.pathmorphing}
\usepackage{tikz}
\usetikzlibrary{shapes.geometric, arrows,patterns,snakes}
\tikzstyle{ellip} = [ellipse, minimum width=3cm, minimum height=1cm,text centered, draw=black]

\newskip\humongous \humongous=0pt plus 1000pt minus 1000pt

\newif\ifdtup

\allowdisplaybreaks[1]

\catcode`\@=11

\@addtoreset{equation}{section}

\def\@normalsize{\@setsize\normalsize{15pt}\xiipt\@xiipt
\abovedisplayskip 14pt plus3pt minus3pt%
\belowdisplayskip \abovedisplayskip
\abovedisplayshortskip \z@ plus3pt%
\belowdisplayshortskip 7pt plus3.5pt minus0pt}

\def\small{\@setsize\small{13.6pt}\xipt\@xipt
\abovedisplayskip 13pt plus3pt minus3pt%
\belowdisplayskip \abovedisplayskip
\abovedisplayshortskip \z@ plus3pt%
\belowdisplayshortskip 7pt plus3.5pt minus0pt
\def\@listi{\parsep 4.5pt plus 2pt minus 1pt
     \itemsep \parsep
     \topsep 9pt plus 3pt minus 3pt}}

\relax

\catcode`@=12

\catcode`\@=11

\def\section{\@startsection{section}{1}{\z@}{3.5ex plus 1ex minus
   .2ex}{2.3ex plus .2ex}{\large\bf}}

\def\SymBoxes#1#2#3#4{\newdimen\un@t \un@t#3%
\raisebox{#1}{\rule{#2\un@t}{#4}\hskip-#2\un@t
\@tempdimb\un@t \advance\@tempdimb by-#4\@tempcntb#2\relax%
\@whilenum{\@tempcntb>0}\do{
\rule{#4}{\un@t}\hskip\@tempdimb \advance\@tempcntb by\m@ne}%
\hskip-#2\un@t \rule[\un@t]{#2\un@t}{#4}%
\rule[\un@t]{#4}{#4}\hskip-#4
\rule{#4}{\un@t}}\hskip-#4}                

\begin{document}

\newcommand{\beq}{\begin{equation}}
\newcommand{\eeq}{\end{equation}}
\newcommand{\bea}{\begin{eqnarray}}
\newcommand{\eea}{\end{eqnarray}}
\newcommand{\beas}{\begin{eqnarray*}}
\newcommand{\eeas}{\end{eqnarray*}}
\newcommand{\defi}{\stackrel{\rm def}{=}}
\newcommand{\non}{\nonumber}
\newcommand{\bquo}{\begin{quote}}
\newcommand{\enqu}{\end{quote}}
\renewcommand{\(}{\begin{equation}}
\renewcommand{\)}{\end{equation}}
\def \eqn#1#2{\begin{equation}#2\label{#1}\end{equation}}
\def\IZ{{\mathbb Z}}
\def\IR{{\mathbb R}}
\def\IC{{\mathbb C}}
\def\IQ{{\mathbb Q}}
\def\de{\partial}
\def\Tr{ \hbox{\rm Tr}}
\def\H{ \hbox{\rm H}}
\def\HE{ \hbox{$\rm H^{even}$}}
\def\HO{ \hbox{$\rm H^{odd}$}}
\def\K{ \hbox{\rm K}}
\def\Im{ \hbox{\rm Im}}
\def\Ker{ \hbox{\rm Ker}}
\def\const{\hbox {\rm const.}}
\def\o{\over}
\def\im{\hbox{\rm Im}}
\def\re{\hbox{\rm Re}}
\def\bra{\langle}\def\ket{\rangle}
\def\Arg{\hbox {\rm Arg}}
\def\Re{\hbox {\rm Re}}
\def\Im{\hbox {\rm Im}}
\def\exo{\hbox {\rm exp}}
\def\diag{\hbox{\rm diag}}
\def\longvert{{\rule[-2mm]{0.1mm}{7mm}}\,}
\def\a{\alpha}
\def\dag{{}^{\dagger}}
\def\tq{{\widetilde q}}
\def\p{{}^{\prime}}
\def\W{W}
\def\N{{\cal N}}
\def\hsp{,\hspace{.7cm}}

\def\br{\nonumber\\}
\def\IZ{{\mathbb Z}}
\def\IR{{\mathbb R}}
\def\IC{{\mathbb C}}
\def\IQ{{\mathbb Q}}
\def\IP{{\mathbb P}}
\def \eqn#1#2{\begin{equation}#2\label{#1}\end{equation}}

\newcommand{\sgm}[1]{\sigma_{#1}}
\newcommand{\idd}{\mathbf{1}}

\newcommand{\C}{\ensuremath{\mathbb C}}
\newcommand{\Z}{\ensuremath{\mathbb Z}}
\newcommand{\R}{\ensuremath{\mathbb R}}
\newcommand{\rp}{\ensuremath{\mathbb {RP}}}
\newcommand{\cp}{\ensuremath{\mathbb {CP}}}
\newcommand{\vac}{\ensuremath{|0\rangle}}
\newcommand{\vact}{\ensuremath{|00\rangle}                    }
\newcommand{\oc}{\ensuremath{\overline{c}}}
\begin{titlepage}
\begin{flushright}
CHEP XXXXX\\
KIAS-P17068
\end{flushright}
\bigskip
\def\thefootnote{\fnsymbol{footnote}}

\begin{center}
{\Large
{\bf Contrasting SYK-like Models
}
}
\end{center}

\bigskip
\begin{center}
{\large  Chethan KRISHNAN$^a$\footnote{\texttt{chethan.krishnan@gmail.com}}, K. V. Pavan KUMAR$^a$\footnote{\texttt{kumar.pavan56@gmail.com}}, and Dario ROSA$^b$\footnote{\texttt{Dario85@kias.re.kr}}\vspace{0.15in} \\ }
\vspace{0.1in}

\end{center}

\renewcommand{\thefootnote}{\arabic{footnote}}

\begin{center}
$^a$ {Center for High Energy Physics,\\
Indian Institute of Science, Bangalore 560012, India\\}
\vspace{0.5cm}
$^b$ {School of Physics, Korea Institute for Advanced Study,\\
Seoul 02455, Korea}

\end{center}

\noindent
\begin{center} {\bf Abstract} \end{center}
We contrast some aspects of various SYK-like models with large-$N$ melonic behavior. First, we note that ungauged tensor models can exhibit symmetry breaking, even though these are 0+1 dimensional theories. Related to this, we show that when gauged, some of them admit no singlets, and are anomalous. The uncolored Majorana tensor model with even $N$ is a simple case where gauge singlets can exist in the spectrum. We outline a strategy for solving for the singlet spectrum, taking advantage of the results in arXiv:1706.05364, and reproduce the singlet states expected in $N=2$.  
In the second part of the paper, we contrast the random matrix aspects of some ungauged tensor models, the original SYK model, and a model due to Gross and Rosenhaus. The latter, even though disorder averaged, shows parallels with the Gurau-Witten model. In particular, the two models fall into identical Andreev ensembles as a function of $N$. In an appendix, we contrast the (expected) spectra of AdS$_2$ quantum gravity, SYK and SYK-like tensor models, and the zeros of the Riemann Zeta function.


\vspace{1.6 cm}
\vfill

\end{titlepage}

\setcounter{footnote}{0}


\tableofcontents

\section{Introduction}

SYK models \cite{coredump} and SYK-like tensor models \cite{Witten, KT-uncolored, tensordump} are quantum mechanical (i. e., 0+1 dimensional) models that could be useful as holographic duals of quantum gravity in AdS$_2$. They both exhibit encouraging features at large-$N$: maximal chaos \cite{MSS}, solvability, and emergent conformal invariance \cite{Polchinski-SYK, Maldacena:2016hyu}. 

However, there are problems regarding both these classes of theories, as fully viable candidates for holography. The disorder averaged models are troublesome in that they are not truly unitary quantum mechanical theories. The tensor models without disorder were proposed \cite{Witten} as a way around this, but it has recently been noted (see \cite{MuruganStanfordWitten} for a brief qualitative discussion of the issue and \cite{KlebanovCount, Minwalla} for detailed computations) that they have way too many extra zero modes due to the presence of the (gauged/ungauged) symmetry in these theories, and this could detabilize the IR fixed point. 

In this paper, we want to further explore some of the differences between the various classes of SYK-like theories that have emerged in the recent past. The results we find include:
\begin{itemize}
\item Some of the tensor models, even though they are quantum mechanical theories in 0+1 dimensions, can exhibit symmetry breaking.
\item A closely related fact is that when gauged, some of these theories are left with no singlet states in the spectrum. We will see that this is a Hamiltonian manifestation of a global anomaly. A simple corollary is that the complexified tensor models do not suffer from any anomalies, and neither do the Majorana tensor models with a suitable even number of fermions, as we will explain. This includes the Gurau-Witten and Klebanov-Tarnopolsky models with even $N$, and is consistent with the fact that their ground states were found to be unique in \cite{KSS, KKS} for low (even) $N$.
\item We will explore the random matrix/quantum chaos behavior of tensor models and SYK-like disorder averaged models, and contrast them in some detail for small $N$. The main outcome of our comparison is to point out and clarify the relationship between the most relevant features (and time scales) of the Spectral Form Factor (SFF) and the correponding hamiltonian spectrum.  More in detail, we will find that the length of the ramp, a non-perturbative effect in SYK-like theories, is controlled by the difference in the mean energy gaps between the tail and the bulk of the spectrum. It would be extremely interesting to see which physical effects determine the differences in the energy gaps.   
\item A specific flavored, disorder-averaged model of Gross-Rosenhaus exhibits close parallels to the Gurau-Witten model. In particular, we will find that the symmetry classification of the random matrix ensembles into which they fall exhibit an identical pattern.
\item In the appendices, we compare and contrast the features of the spectrum of AdS$_2$ quantum gravity, SYK and SYK-like models and (for amusement!) the Hamiltonian that is expected to reproduce the zeros of the Riemann zeta function. We find many parallels, but also some important differences. Finally, we re-analyze the first hints of level repulsion for the GW model, that have been found in \cite{KKS} using the unfolded level spacing distribution, using the diagnostic of the so-called $r$-statistics. We will see that the results of the $r$-statistics do {\it not} agree with the results of the unfolded level spacing distribution, and we will discuss an explanation of this discrepancy.
\end{itemize}

It is an interesting question whether one can reproduce all the good features of perturbative large-$N$ melonic models, without having to sacrifice unitarity via a disorder average or by  introducing unwanted destabilizing zero  modes. It seems that what one needs is the ability to get large-$N$ melonic behavior with neither a disorder average nor extra symmetries whose rank is $N$, if one wants to reproduce all the features of dilatonic AdS$_2$ quantum gravity. Is it possible to construct such a theory?

In some ways, the extra zero modes found in tensor models are analogous to the extra light states found in the minimal model holography of \cite{MinimalModelHolographyReview}. Unfortunately, we will not have much to say about these very interesting questions in this paper. 


\section{Symmetry Breaking, Gauge Singlets and Anomalies}

In this section we will discuss some general aspects of tensor models. Our observations are simple, but we suspect they are not as widely appreciated as they should be. Our discussion in many parts is a Hamiltonian analogue of the closely related work of \cite{Schwimmer}\footnote{We thank I. Klebanov, S. Minwalla and E. Witten for helpful discussions/correspondence on these matters. As this paper was being finalized, revised versions of \cite{KlebanovCount, Minwalla} have appeared, which make comments related to some of the points we make in this section.}. 

We will discuss the simplest tensor models that have melonic large-$N$ behavior: namely the Majorana fermion theory with $O(N)^3$ symmetry group written down by Klebanov-Tarnopolsky \cite{KT-uncolored} (see also \cite{Tanasa}). This theory is defined by the action
\bea
S=\int dt ~\left(\frac{i}{2}\psi ^{ijk}\partial _t\psi ^{ijk}+\frac{1}{4}\psi ^{ijk}\psi ^{ilm}\psi ^{pjm}\psi ^{plk}\right)  \label{KTint}
\eea

In \cite{KKS} the simplest ungauged models of this type were diagonalized via brute force on a computer: the cases with $N=2, 3$. In \cite{finite-N} it was shown that for even $N$ there exists an approach for diagonalizing these theories which should be tractable on a computer for small enough values of $N$. However a discussion of the gauge singlet spectrum was not undertaken there. 

From a holographic perspective, it is natural to expect that the (singlet sector of the) gauged models are the most interesting.  One reason for this is that holography is an aspect of open-closed duality, and therefore it would be encouraging to find an analogue of open string gauge invariance in the boundary theory. Another more recently suggested reason \cite{PolchErrorCorr} to expect boundary gauge invariance is related to bulk locality and the quantum error correction aspect of the hologram \cite{AlmDongHarlow}\footnote{We note however that the arguments there strictly only apply to higher dimensional holography, but they are nontheless suggestive.}. In any event, we will present a discussion of some aspects of the gauged models in this section.

Before addressing the question of singlets however, we will first consider the question of symmetry breaking in the ungauged theory, which is closely related.

\subsection{Symmetry Breaking and Charge Singlets}

It is an oft-stated truism that symmetry breaking cannot occur in 0+1 dimensions, aka quantum mechanics. It is less often stated, but implicitly well-known to everyone (as we will now demonstrate), that this truism cannot be right. 

The trouble with this statement is that it tacitly assumes {\em bosonic} quantum mechanics. The usual argument against such bosonic symmetry breaking relies on potentials (ie., functions of bosonic variables like $x$) with degenerate minima. In 0+1 dimensions, these have non-zero amplitudes for tunneling between them, leading eventually to a unique ground state and no symmetry breaking. In higher dimensions on the other hand, the infinite spatial volume suppresses tunneling between vacua, and leads to the fact that broken symmetries do exist in quantum field theory.

However, it is trivial to construct symmetry breaking 0+1 d quantum mechanical systems if one uses fermionic variables. Consider the free theory of $N$ real (Majorana) fermions with the Lagrangian
\bea
L=\frac{i}{2} \psi^i \partial_t \psi^i.
\eea
The canonical anti-commutators force $\psi$'s to satisfy the Clifford algebra. Since the Hamiltonian arising from this Lagrangian is identically zero, all states in the Hilbert space are eigenstates with degenerate energy, leading to symmetry breaking. Note that an analogous construction for a bosonic theory with $L= \frac{1}{2} (\partial_t x)^2$ fails to produce symmetry breaking because the Hamiltonian is non-zero, and the zero momentum state is the unique ground state. 

Since degeneracy means that the degenerate levels form a representation of some symmetry, and since all states above are manifestly degenerate in energy, this is enough to exhibit symmetry breaking in this simple quantum mechanial theory\footnote{We note however that in the case when $N$ is even, this representation is reducible: there is a discrete symmetry which can be used to further split the Hilbert space into two halves with distinct ($\pm$) eigenvalues. This is just the familiar fact that even dimensional Dirac spinors admit a breakup into Weyl and anti-Weyl representations. In the case $N=2$, this leads to a subtlety: there are only two states in the Hilbert space, and the Weyl spinors in two dimensions are one dimensional and automatically singlets of $SO(2) \sim U(1)$. We will discuss this from a different angle when we discuss global gauge anomalies.}. But for our future goals, it is useful to explicitly demonstrate that there are no states in the Hilbert space that are charge singlets under the symmetry group of the theory, which is $O(N)$. The Noether charges are trivial to compute. They take the form
\bea
Q^{ij}=i \psi^i \psi^j
\eea
with $i < j$. A singlet state is one that is annihilated by all of these charges. When $N$ is even, it is convenient to split the $\psi^i$ into creation and anniilation operators in the Clifford algebra as:
\begin{align}
\psi ^{i^\pm}&=\frac{1}{\sqrt{2}}\left(\psi ^i\pm i \psi ^{i+1}\right)
\end{align}
where $i^{\pm}$ takes values from 1 to $\frac{N}{2}$ and is related to (odd) $i$ as:
\begin{align}
i&=2i^{\pm}-1
\end{align} 
In terms of these creation and annihilation operators, for $i\neq j-1$, the Noether charge $Q^{ij}$ has four different forms depending on whether each of $(i,j)$ is even or odd. These four different forms can be written as:
\begin{align}
Q^{ij}&=\frac{i}{2}\left(\psi ^{i^+}\psi ^{j^+}+\psi ^{i^-}\psi ^{j^-}+\psi ^{i^+}\psi ^{j^-}+\psi ^{i^-}\psi ^{j^+}\right)\nonumber \\
Q^{(i+1)(j+1)}&=\frac{-i}{2}\left(\psi ^{i^+}\psi ^{j^+}+\psi ^{i^-}\psi ^{j^-}-\psi ^{i^+}\psi ^{j^-}-\psi ^{i^-}\psi ^{j^+}\right)\nonumber \\
Q^{i(j+1)}&=\frac{1}{2}\left(\psi ^{i^+}\psi ^{j^+}-\psi ^{i^-}\psi ^{j^-}-\psi ^{i^+}\psi ^{j^-}+\psi ^{i^-}\psi ^{j^+}\right)\nonumber \\
Q^{(i+1)j}&=\frac{1}{2}\left(\psi ^{i^+}\psi ^{j^+}-\psi ^{i^-}\psi ^{j^-}+\psi ^{i^+}\psi ^{j^-}-\psi ^{i^-}\psi ^{j^+}\right)
\end{align}
where $i$ and $j$ are now both odd. We can now take a linear combination of these charges and define a new set of charges as:
\begin{align}
Q_1^{ij}&=\psi ^{i^+}\psi ^{j^+}\nonumber \\
Q_2^{ij}&=\psi ^{i^-}\psi ^{j^-}\nonumber \\
Q_3^{ij}&=\psi ^{i^+}\psi ^{j^-}\nonumber \\
Q_4^{ij}&=\psi ^{i^-}\psi ^{j^+}
\end{align} 
Note that these new charges are defined only when $i\neq j- 1$. For the case of $i=j- 1$, we have only one charge and is given by:
\begin{align}
Q_5^{i(i+1)}&=\psi ^{i^+}\psi ^{i^-}- \psi ^{i^-}\psi ^{i^+}=2\psi ^{i^+}\psi ^{i^-}-1
\end{align}
where $i$ is odd and there is no summation over $i$.

The singlet states are the states that are annihilated by all these charges i.e.,
\begin{align}
Q_a|\text{singlet}\rangle &=0
\end{align}
Starting from the $Q_5$ charge, we can show that all the singlet states are at $\frac{N}{4}$th Clifford level. Noting that the total Clifford levels are $\frac{N}{2}$ in number, we can conclude that all the singlet states are at mid Clifford level if $N$ is a multiple of 4 and there are no singlet states in the spectrum if $N ~\text{mod}~ 4=2$\footnote{Note however that in the $N=2$ case, when we gauge the theory, we must break charge conjugation to save the gauge invariance. So we will have gauge singlets in that case. Here we are talking about absence of global, not gauge, singlets.}.

To understand the action of $Q_1$ charge, we first note that a possible singlet state is given by:
\begin{align}
\sum \alpha _{i_1^+\ldots i_{N/4}^+}~\psi ^{i_1^+}\ldots \psi ^{i_{N/4}^+}|~\rangle
\end{align} 
where $|~\rangle $ is the Clifford vacuum and $\alpha _{i_1^+\ldots i_{N/4}^+}$ are numerical coefficients. For the charge $Q_1$ to annihilate this state, all the $\alpha $'s need to be zero\footnote{Note that $N$ is a multiple of 4. When $N \ge 8$, one can argue straightforwardly that just the $Q_1$ charge condition is enough to rule out charge singlets. For $N=4$ however, the $Q_1, Q_2$ charges yield no constraints. But an explicit calculation shows that $Q_3$ and $Q_4$ do, and that they are enough to rule out singlets.}. This implies that there are \textit{no} singlet states in a free theory of even number of $O(N)$ Majorana fermions. 

For the case of odd $N$, we can work\footnote{Note that the Clifford algebra determines the dimensionality of the Hilbert space. For the odd $N$ case, the Hilbert space dimension is $2^{(N-1)/2}$, which is same as that of $(N-1)$ case. Thus, the Hilbert space can be completely constructed by the creation and annihilation operators of first $(N-1)$ Majorana fermions.} with the creation and annihilation operators constructed for the first $(N-1)$ Majorana fermions. The arguments we made in the even $N$ case go through here as well and it can be shown that there are no singlet states in the odd $N$ case too. To summarize, there are no singlet states present in a free theory of $O(N)$ Majorana fermions. 

When the theory has a non-trivial interaction term, say of the form \eqref{KTint}, the above discussion needs modification, but only in detail. The structure of the Hilbert space as a Clifford representation remains intact, but now since the Hamiltonian is non-trivial, the degeneracies of the eigenstates depend on the energy. To decide whether there is symmetry breaking, one needs to understand whether the ground state is degenerate or unique. If the ground state is degenerate, some symmetry is broken. In the case of the $N=2$ theory, we found in \cite{KKS} that the ground state is unique, suggesting that there could be singlets in the spectrum. The $N=3$ case was degenerate, and the operators that lead to the degeneracies there were identified in \cite{KKS}. Thus this case exhibits symmetry breaking. The symmetry operators constructed in \cite{KKS} can be generalized to other odd $N$, so we expect that all the odd $N$ cases exhibit symmetry breaking and ground state degeneracy. Note also that the degeneracies discovered in \cite{KKS} affect all levels and not just the ground state. Closely parallel statements should be possible for the Gurau-Witten model, and the results found in \cite{KSS} are consistent with these expectations. 

This leads us to the next question, which is about gauge singlet states in the spectrum. We will now relate the above symmetry breaking phenomena in the ungauged theory to the existence of global gauge anomalies in the corresponding gauged theory.  

\subsection{Anomalies in Quantum Mechanics}

The existance of symmetry breaking in the theories we discussed in the last subsection are intimately connected to the existance of global gauge anomalies in the theory \cite{Schwimmer}. The latter manifests itself in the Hamiltonian/spectrum language via the fact that all the states are non-trivially charged, and therefore when we gauge the theory, the entire Hilbert space has to be thrown away, leaving no theory. 

That anomalies can exist in (Majorana) fermionic quantum mechanics has been noted before using a more conventional path-integral based approach \cite{Schwimmer}. The anomaly is a global gauge anomaly that shows up as a sign in the path integral. For the gauge group $SO(N)$ for all values of $N >2$  there is an anomaly. The $N=2$ case can be re-interpreted as a complex fermion with $U(1)$ charge and there is no anomaly. This was discussed from the Hamiltonian language in footnote 3. 

For odd $N$ the fact that the anomaly is a sign, necessarily implies that there is always bound to be an anomaly in the $O(N)^3$ case in our tensor models. In the even $N$ case, there is no sign anomaly in the $O(N)^3$ case, so singlets can exist in the spectrum. This is related to  the fact that for each $O(N)$ the other $O(N)$'s are flavor indices and since there are an even number of them, the sign anomaly can be avoided\footnote{We thank I. Klebanov and E. Witten for clarifications on this point.}. 

A warm up exercise towards the full SYK-like tensor models is to work with the free $O(N)^2$ theory and try to identify singlet states in its Hilbert space:
\bea
L=\psi^{ij} \partial_t \psi^{ij}.
\eea
One can show by (more elaborate) methods very similar to the ones we used in the previous subsection that this theory, when gauged, has no singlets in its spectrum and is anomalous for $N=4$. We have not checked the status of singlets for arbitrary (even) $N$ in this theory. Instead of looking at these toy models in detail here, we will do a similar calculation in the $O(2)^3$ uncolored tensor model in the next subsection. More discussions on the finite-$N$ singlet spectrum of some classes of tensor models will be presented in a forthcoming paper, where a complete analytic solution of the smallest non-trivial gauged Gurau-Witten model (whose eigenvalue spectrum was found numerically in the ungauged case in \cite{KSS}) can also be found \cite{upcoming}. 

\subsection{Singlet Spectrum of $O(2)^3$ theory}

In this section, we find the spectrum of $O(2)^3$ theory using the technology that is developed in \cite{finite-N}. The construction of the spectrum is based on the fact that Clifford representation forms a basis of the Hilbert space. Further, we identify the singlets in this case and show that they match the singlets that were found in \cite{Loga} in a different language. 

We first review the relevant parts of \cite{finite-N}. We start by noting that the fermionic tensors $\psi ^{ijk}$ obey the following anti-commutation relations:
\begin{align}
\{\psi ^{ijk},\psi ^{pqr}\}&=\delta ^{ip}\delta ^{jq}\delta ^{kr}
\end{align}
Following \cite{finite-N}, we define the following creation and annihilation operators:
\begin{align}
\psi ^{ijk^{\pm}}&=\frac{1}{\sqrt{2}}\left(\psi ^{ijk}\pm i \psi ^{ij(k+1)}\right)
\end{align}
where $k^{\pm}$ takes values\footnote{We are using the notation that we are working with the $O(n)^3$ theory with $n=2$, in this subsection.} from 1 to $\frac{n}{2}$ and  are related to $k$ as follows:
\begin{align}
k&=2k^{\pm}+1
\end{align}
We now define the Clifford vacuum as the state that is annihilated by all the annihilation operators:
\begin{align}
\psi ^{ijk^-}|~\rangle &=0
\end{align}
We can now construct the entire Clifford representation by acting on the ground state with the creation operators. That is, a general state at level $r$ can be constructed as follows:
\begin{align}
\psi ^{i_1j_1k^+_1}\psi ^{i_2j_2k^+_2}\ldots \psi ^{i_rj_rk^+_r}|~\rangle
\end{align}  
Noting that there are $\frac{n^3}{2}$ total creation operators, we see that there are ${n^3/2}\choose {r}$ states at level $r$. Also, because of the anticommutation relations of $\psi ^{ijk^+}$'s, we can show that there are only $\frac{n^3}{2}$ levels. So, total number of states is $2^{n^3/2}$, which is the dimensionality of the Hamiltonian.

We define the level operator as follows:
\begin{align}
Q&=\sum \psi ^{ijk^+}\psi ^{ijk^-}
\end{align}
It can be shown that the Hamiltonian commutes with the level operator. Hence the eigenstates of the Hamiltonian are linear combination of the states that are at same level. Essentially, this allows us to find the eigenstates at each level separately.

We now proceed to find the eigenstates as follows. To start with, we note that any state level $r$  in Clifford representation is obtained by taking tensor product of $r$ fundamental representations of $O(n)\times O(n)\times SU(n/2)$. That is, in the Young tableaux language, a level $r$ state can be written as:
\begin{align}
\left(~\begin{ytableau}
i_1 &\none[\otimes] &i_2 &\none[\otimes]&\none[\dots]&\none[\otimes] &i_r 
\end{ytableau}~ ,
~\begin{ytableau}
j_1 &\none[\otimes] &j_2 &\none[\otimes]&\none[\dots]&\none[\otimes] &j_r
\end{ytableau}~,~ \begin{ytableau}
k^+_1 &\none[\otimes] &k^+_2 &\none[\otimes]&\none[\dots]&\none[\otimes] &k^+_r
\end{ytableau}~ \right)
\end{align} 
Now, the action of Hamiltonian on a general state at level $r$ is given by:
\begin{align}
&\left(H-\frac{n^4}{4}\right)\left(~\begin{ytableau}
i_1 &\none[\otimes]&\none[\dots]&\none[\otimes] &i_r 
\end{ytableau}~ \right.,
\left.~\begin{ytableau}
j_1 &\none[\otimes] &\none[\dots]&\none[\otimes] &j_r
\end{ytableau}~,~ \begin{ytableau}
k^+_1 &\none[\otimes] &\none[\dots]&\none[\otimes] &k^+_r
\end{ytableau}~  \right) \nonumber \\
= & 4n\sum _{p<q} (-1)^{p+q-1} \left[\left(\bullet _{i_pi_q}~,~ \begin{ytableau}j_q& \none[\otimes]&j_p \end{ytableau}~,~k^+_p,k^+_q\right)-\left(\begin{ytableau}i_q&\none[\otimes]&i_p \end{ytableau}~,~ \bullet _{j_pj_q}~,~ \begin{ytableau}k_p^+&\none[\otimes]&k_q^+ \end{ytableau}\right)\right]  \nonumber \\
&\hspace{22 mm}\otimes\left(	\underbrace{~\begin{ytableau}
i_1 &\none[\otimes] &\none[\dots]&\none[\otimes] &i_r
\end{ytableau}~}_{\text{no} ~i_p ~\& ~i_q}~,~ 
\underbrace{~\begin{ytableau}
j_1 &\none[\otimes] &\none[\dots]&\none[\otimes] &j_r
\end{ytableau}~ }_{\text{no} ~j_p ~\& ~j_q}~,~ \underbrace{\begin{ytableau}
k_1^+ &\none[\otimes] &\none[\dots]&\none[\otimes] &k_r^+
\end{ytableau}}_{\text{no} ~k^+_p ~\& ~k^+_q}\right)
\end{align} 
Since the Hamiltonian is a singlet operator under $O(n)\times O(n)\times SU(n/2)$, its eigenstates can be obtained by comparing the irreps of $O(n)\times O(n)\times SU(n/2)$ on both sides. Also, the eigenvalues can be obtained from the knowledge of Clebsch-Gordan coefficients.

From now on, we specialize to $n=2$ case. There are 16 states in the spectrum of which 14 states have zero\footnote{Note that we have shifted the entire spectrum by a constant} energy. The remaining two states are given by:
\begin{align}
\label{singlet-1,n=2}
\frac{1}{2}\epsilon _{i_1i_2}\delta _{j_1j_2}\psi ^{i_1j_11^+}\psi ^{i_2j_21^+}|~\rangle=\left(\psi ^{111^+}\psi ^{211^+}+\psi ^{121^+}\psi ^{221^+}\right)|~\rangle \\
\label{singlet-2,n=2}
\frac{1}{2}\delta _{i_1i_2}\epsilon _{j_1j_2}\psi ^{i_1j_11^+}\psi ^{i_2j_21^+}|~\rangle=\left(\psi ^{111^+}\psi ^{121^+}+\psi ^{211^+}\psi ^{221^+}\right)|~\rangle
\end{align}
The energies of these states are `$+8$' and `$-8$' respectively. That is, the $n=2$ spectrum has 14 states at mid-level and the other two states are located equidistant from $E=0$. This is the exact structure that is found by numerically diagonalizing the $n=2$ theory. 

We now proceed to find the singlet states in the $n=2$ theory. Before doing that, we recall ourselves that singlet states have a zero charge under each of the three $O(n)$'s. The charge operators themselves can be computed using Noether's theorem and are given by:
\begin{align}
\label{O(n) charges}
Q_1^{i_1i_2}&=i ~\psi ^{i_1jk}\psi ^{i_2jk} \nonumber \\
Q_2^{j_1j_2}&=i ~\psi ^{ij_1k}\psi ^{ij_2k} \nonumber \\
Q_3^{k_1k_2}&=i ~\psi ^{ijk_1}\psi ^{ijk_2} 
\end{align} 
All the singlet states should have zero charge under these three operators i.e., 
\begin{align}
Q_a |\text{singlet}\rangle &=0
\end{align}
where $a=\{1,2,3\}$. Note that this is a necessary and sufficient condition for a state to be a singlet under $O(n)^3$. 

For the case of $n=2$, the charges \eqref{O(n) charges} can be written as follows:
\begin{align}
Q_1^{12}&=i \left(\psi ^{111^+}\psi ^{211^-}+\psi ^{111^-}\psi ^{211^+}+\psi ^{121^+}\psi ^{221^-}+\psi ^{121^-}\psi ^{221^+}\right)\nonumber \\
Q_2^{12}&=i \left(\psi ^{111^+}\psi ^{121^-}+\psi ^{111^-}\psi ^{121^+}+\psi ^{211^+}\psi ^{221^-}+\psi ^{211^-}\psi ^{221^+}\right)\nonumber \\
Q_3^{12}&=2-\psi ^{111^+}\psi ^{111^-}-\psi ^{121^+}\psi ^{121^-}-\psi ^{211^+}\psi ^{211^-}-\psi ^{221^+}\psi ^{221^-}
\end{align}
Our goal is to find singlet states among the energy eigenstates. To do so, we need to identify the states among the energy eigenstates that are annihilated by all of these charges. Note that only the states at second Clifford level are annihilated by $Q_3^{12}$. Out of the six eigenstates at level-2, only two of them are annihilated by both $Q_1$ and $Q_2$ and are given by:
\begin{align}
\left(\psi ^{111^+}\psi ^{211^+}+\psi ^{121^+}\psi ^{221^+}\right)|~\rangle \\
\left(\psi ^{111^+}\psi ^{121^+}+\psi ^{211^+}\psi ^{221^+}\right)|~\rangle
\end{align}
So, we have shown explicitly that the non-zero energy eigenstates are the singlets in case of $n=2$ theory. These findings are consistent with that of \cite{Loga}. It is straightforward to see by adapting the arguments in this section, that for generic even $N$, the singlet states lie in the mid-Clifford level. We will not elaborate on this further here, a more detailed discussion will be presented elsewhere.

We conclude this section with a speculative comment: Since the consitency of these theories relies on the absence of a $\IZ_2$  anomaly, it is tempting to think that this is a hint of the dual AdS$_2$ having {\em two}  boundaries. The singlets in the spectrum should perhaps be understood in terms of Wilson lines stretching from boundary to boundary? Could one construct a Maldacena-like eternal black hole \cite{MaldaEternal} where instead of working with singlet states separately from both boundary, one should construct singlets in the doubled tensor product Hilbert space? Does this philosophy have applications more generally than in 0+1 dimensional holography? We will not discuss these speculations further in this paper.

\section{Contrasting the RMT features of the GW model and of SYK}

We now move to the second part of the paper. Our aim is to study some spectral properties of the GW model and of SYK. Spectral studies for both the models have been performed in the literature, \cite{Cotler} \cite{KSS}. However, our aim will be different: we will connect some peculiar features of the Spectral Form Factors (SFFs) with some peculiarities of the spectra. We will see that some quantitative predictions on the behavior of the SFF can be made by some simple analysis of the eigenvalues distribution of the hamiltonian. More in details, we will see that the length of the ``ramp'' (a distinctive feature of random matrix theories (RMT)) is in direct relation with the different mean energy gaps computed in the tail and in the bulk of the eigenvalues spectra. We will only work with the ungauged tensor model in this section.

\subsection{A brief review of the GW model}
Let us review the main features of the GW model. The GW model is a particular instance of  ``colored'' tensor model quantum mechanics. As such, the system is defined in $0 + 1$ dimensions and it contains $N = (D+1) \, n^D$ fermions, collectively called $\psi_\mu^{i_1 i_2 \dots i_D}$. Each of the indices $i_1 , \, i_2 \, \dots$ are $O(n)$ indices; the index $\mu$ (the color index) runs from $0$ to $D$. We will consider the case with $n = 2$ and $D = 3$ hence our fermions are of the form $\psi_\mu^{i j k}$, with $\mu = 0, \cdots , 3$ and $i$, $j$, $k$ being $O(2)$ indices. The Hamiltonian for this particular case takes the form
\begin{align}
\label{eq:GWhamiltonianspecific}
H = \frac{J}{\sqrt{8}}  \sum_{i,j,k,l,m,n}  \, \psi_0^{ijk} \psi_1^{i m l} \psi_2^{n j l} \psi_3^{n m k}  \ .
\end{align}

Let us recall also the main features of SYK. It describes a set of  N Majorana fermions $\psi_a$ ($a= 1 , \cdots , N$) in $0 + 1$ dimensions. The Hamiltonian takes the form
\begin{align}\label{eq:hamiltonianSYKgeneral}
H = \frac{1}{4!} \, \sum_{a,b,c,d} J_{abcd} \, \psi_a \psi_b \psi_c \psi_d \ ,
\end{align}
where $J_{abcd}$ is a completely anti-symmetric tensor, and each element is a random real number extracted from a Gaussian distribution with zero mean value and variance $\sigma^2 = \frac{3!}{N^3} J^2$, with $J$ a fixed number that we will set equal to $1$. 

The GW model, contrary to SYK, has one coupling constant $J$ and there is no disorder to average. Hence, it is a much more traditional example of QM. Its large $N$ expansion is dominated by a set of diagrams which are called {\it melonic} and which coincide with the dominant diagrams in SYK {\it after} the average procedure. Therefore, the GW model (and, more generally, many tensor models) share with SYK its nice perturbative large-$N$ features, without the need of introducing disorder averaging.

The last feature we want to recall is the presence of the so-called {\it spectral mirror symmetry}: the GW theory has an operator $S$ which is unitary and that {\it anti-commutes} with the Hamiltonian. Hence, the eigenvalues of the Hamiltonian come in pairs $\pm E_i$. From the RMT point of view, the spectral mirror symmetry implies that the GW model should not be described by the standard random matrices ensembles (GUE, GOE and GSE) but by the other, non-standard, ensembles of the Altland-Zirnbauer classification. A precise analysis of the symmetry classes for the GW model, for all the values of $D$ and $n$, can be found in \cite{KKS}.

We will consider in this Section the case of SYK with $N = 32$ and GW with $n=2, D=3$ which also leads to $N=32$. However, even though we consider both the models with the same number of fermions, we caution the reader that the direct comparison between the two models is not automatic. This is because in the large-$(N,n)$ limit, the $J$'s in the two theories scale differently. SYK has a natural $N^{3/2}$ scaling whereas GW has a natural $n^{3/2}=\frac{\sqrt{N}}{2}$ scaling. Indeed, the large $N$ limits of the two theories are different, starting at subleading orders \cite{Witten}. So even though we work with the same value of $N=32$ for SYK and GW, we emphasize that we are not implying a strict parallel between the corresponding models at this value of $N$. The case we consider is the smallest non-trivial GW model, because it also happens to be the biggest GW model we could diagonalize: it is quite plausible that the RMT features will be more robust for larger-$N$. Our main goal in this section is not to make an actual comparison between the two models, but rather to understand how the properties of the spectrum affect the SFF.

\subsection{The spectral form factor}\label{sec:SFF}

We are going to investigate the {\it Spectral Form Factor} (SFF). It is defined by the expression 
\begin{align}\label{eq:SFFdefinition}
& g (t; \,\beta) \equiv \frac{|Z(t; \, \beta)|^2}{|Z(0;\,\beta)|^2} \ , \qquad Z (t ; \, \beta) \equiv \Tr \, (e^{- (\beta + i t) \, H}) \ ,
\end{align}
and it is known to be a probe of the chaotic properties of a model.

We have diagonalized numerically the Hamiltonian (\ref{eq:GWhamiltonianspecific}) and computed the spectral form factor using (\ref{eq:SFFdefinition}) for various values of $\beta$. 
The plots for SYK and for the GW model have similarities, but also differences. The SFF for SYK displays an initial decay, until a time of order $t \, {\sim} \,  5 \cdot 10^2$,  that carries the value of the SFF to values which are very small (around $10^{-7}$ for small values of $\beta$ and $10^{-4}$ for large values of $\beta$).  After this time (the ``dip'' time) the SFF grows linearly for a long period of time, until it reaches a plateau. The value of the plateau is still exponentially suppressed, even for large values of $\beta$. After the dip the SFF oscillates erratically \cite{Cotler}. 

\begin{figure}
\centering
    \subfigure{\includegraphics[width=0.6\textwidth]{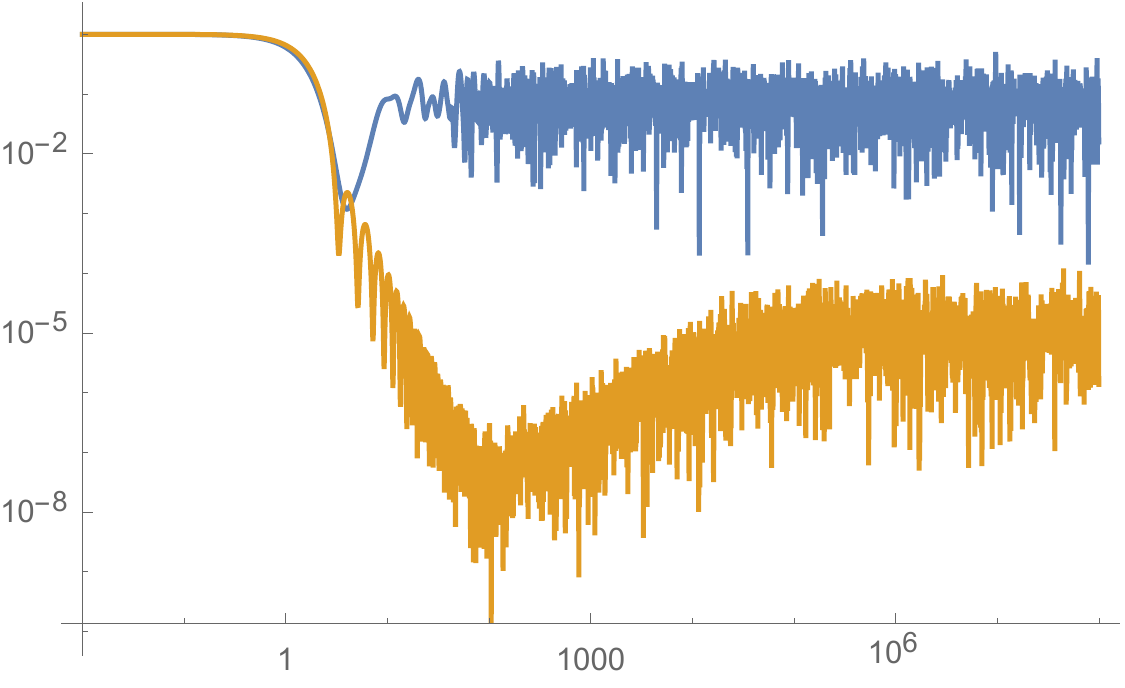}}
    \hspace{.4cm}
   \subfigure{\includegraphics[width=0.6\textwidth]{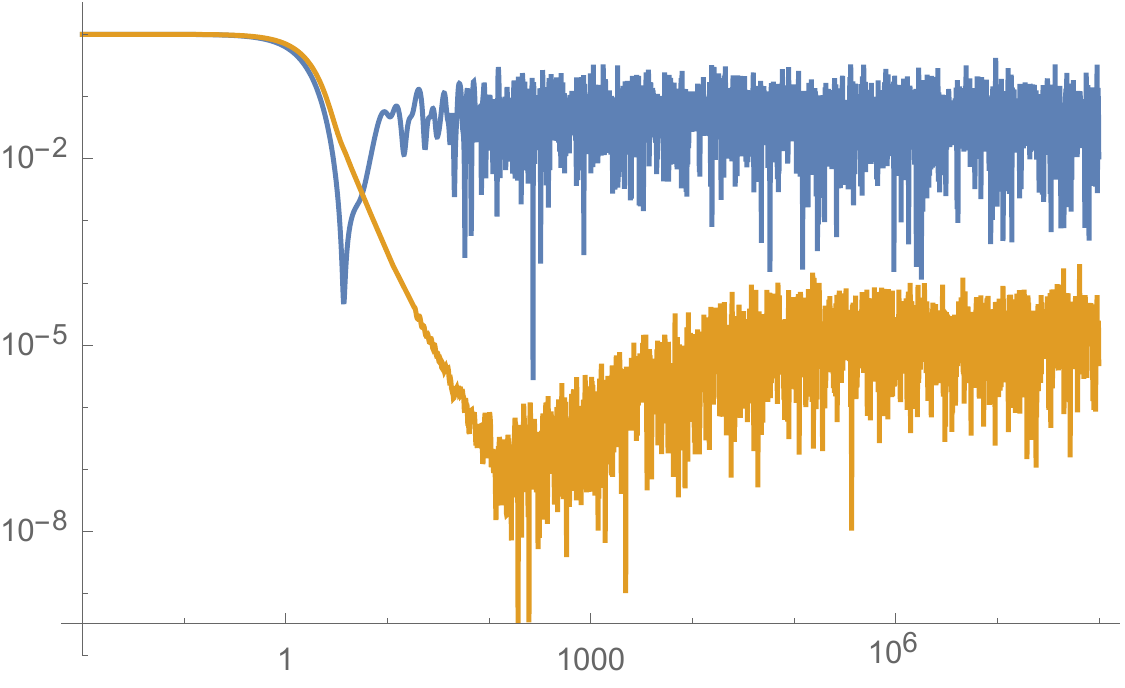}}
    \hspace{.4cm}
   \subfigure{\includegraphics[width=0.6\textwidth]{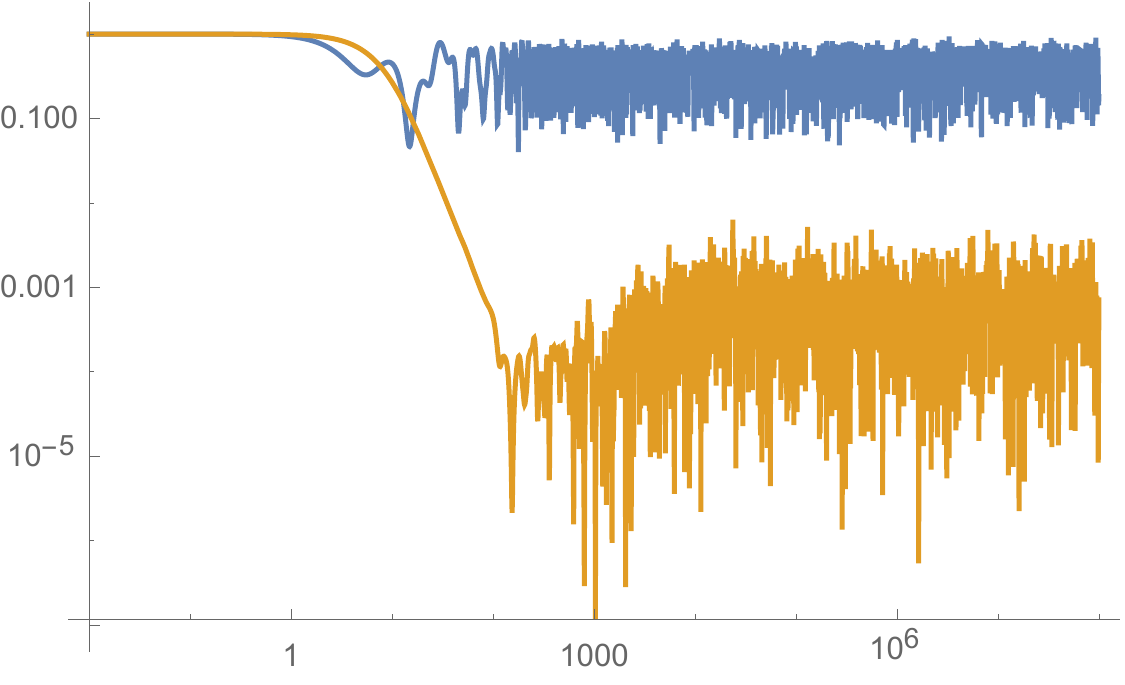}}
   \caption{The SFF for the GW model (blue color) and for SYK (yellow color). Top: $\beta = 0.1$, center: $\beta = 1$, bottom: $\beta = 10$. The GW model is $n=2, D=3$, the SYK model is $N=32$.}
    \label{fig:comparison_SYK_GW_beta01}
\end{figure}

In the GW model, for small values of $\beta$ ($\beta = 0.1$ and $\beta = 1$) we continue to see  a pattern qualitatively similar to what we observed for SYK, with an initial decay followed by a ramp and a plateau characterized by erratic fluctuations of the SFF. However the similarities stop here: the initial decay is much more rapid and the dip time comes much earlier (around $t \sim 1$). Also the ramp  is much more ripid than in SYK and the height of the plateau is much higher. The differences become even more dramatic for larger values of $\beta$ and for $\beta = 10$ the SFF loses most of its structure: the slope, ramp and plateau all become indistinct. Finally, in all the plots we see that the SFF at the dip time is much bigger in the GW model (around $10^{-2}$). The plots for SYK and GW model together are presented in Figure \ref{fig:comparison_SYK_GW_beta01}.


We will explain these  differences in terms of the differences in spectra between GW and SYK. Before doing that, we discuss how we can properly cure the erratic behavior at late times.

\subsubsection{Removing the erratic behavior: the progressive time average}

A typical feature of the SFF is the erratic behavior for late times: these rapid oscillations for large values of $t$ make hard to extract the relevant information from the plot with sufficient precision. This problem, in the context of RMTs, is cured by performing an {\it ensemble average}: one considers many realizations of the sample and then performs an average over these realizations.  The same procedure can be performed for SYK by taking various realizations of the coupling constants and then averaging over them. 

In the case of tensor models the situation is more involved: there is no natural process of averaging at disposal. The authors of \cite{Cotler} suggested to perform a {\it time averaging} procedure: for each time $t$ one considers a {\it fixed} temporal window, of width $\Delta \, t$ suitably chosen, and then performs the average over that window. The size of $\Delta \, t$ has to be chosen with care: if it is too small, the average will be not strong enough to wash out the erratic behavior; if it is too big all the ``slope - ramp'' behavior will be lost. In the paper \cite{Balasubramanian:2016ids} it has been argued (in a different context) that this tension cannot be solved: it is not possible to find an optimal windows which preserves both the early time ``slope-ramp'' behavior {\it and} that removes the erratic behavior. This conclusion has been confirmed in the GW model by \cite{KSS}.  

The authors of \cite{Balasubramanian:2016ids}  have also proposed an alternative way to perform the time average. The idea is very simple: since in the early time a small time window is necessary to correctly reproduce the ``slope-ramp'' feature, whereas in the late time we need a sufficiently big time window to make smooth the erratic behavior; they have proposed a {\it progressive time average procedure}, in which the width $\Delta \, t$  is proportional to the time $t$. 

\begin{figure}
\centering
    \subfigure{\includegraphics[width=0.6\textwidth]{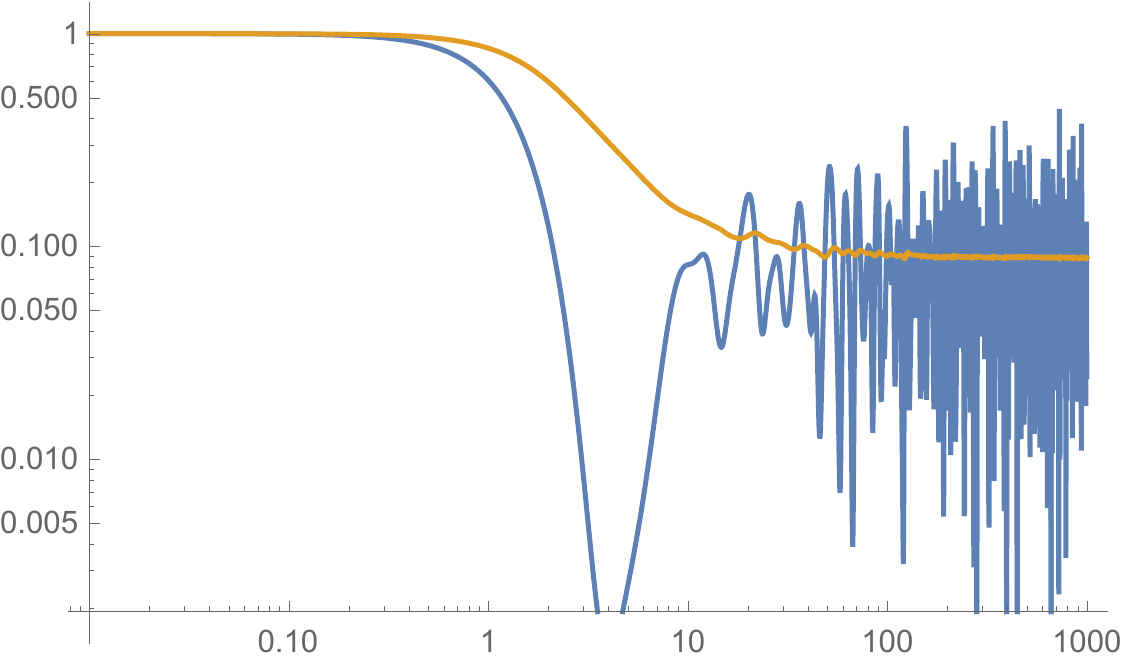}}
    \hspace{.4cm}
   \subfigure{\includegraphics[width=0.6\textwidth]{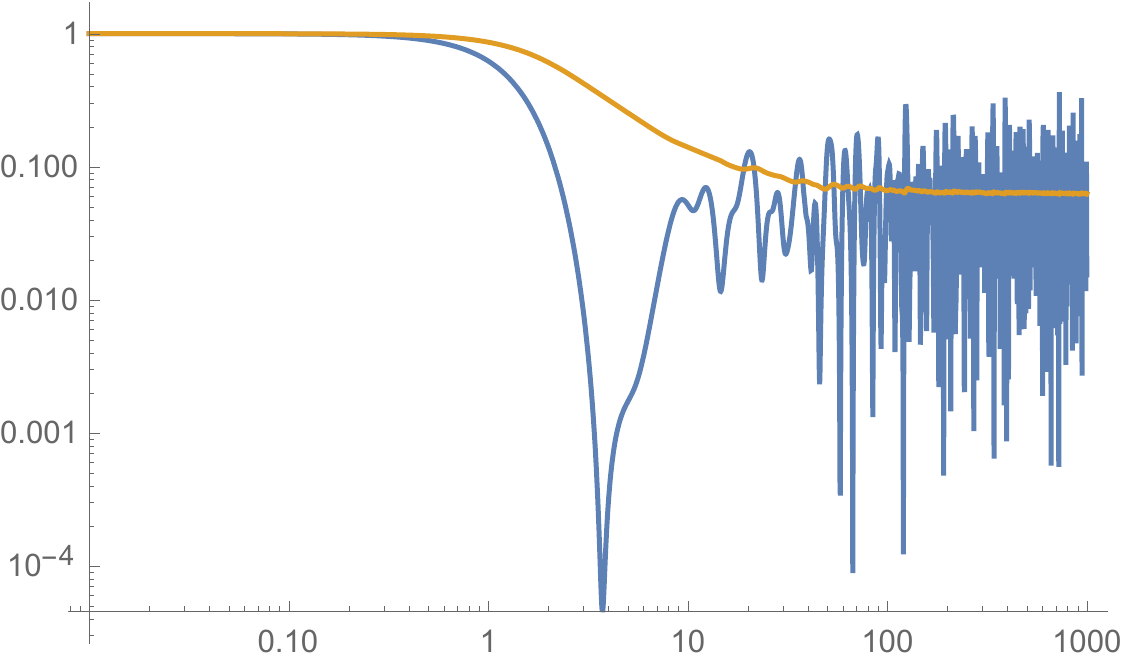}}
    \hspace{.4cm}
   \subfigure{\includegraphics[width=0.6\textwidth]{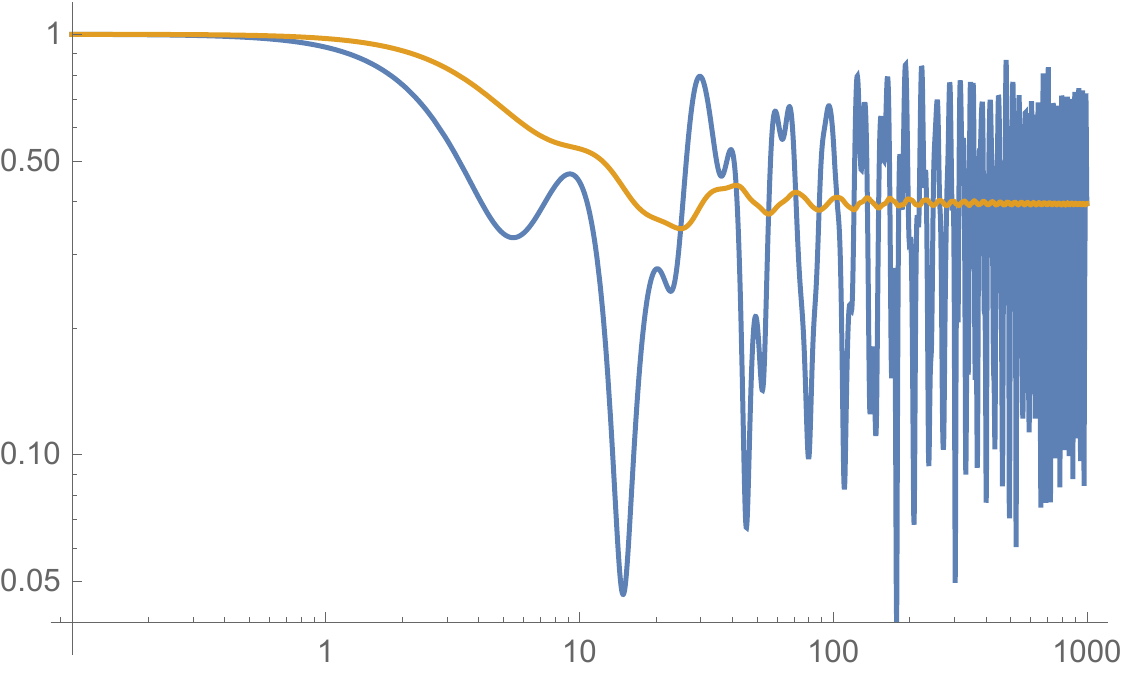}}
   \caption{The progressive time average (\ref{eq:PTA_naive_definition}) for the GW model. Top: $\beta = 0.1$, center: $\beta = 1$, bottom: $\beta = 10$}
    \label{fig:naive_PTA_GW_beta01}
\end{figure}

We want to show now that the progressive time average works well also for the GW model (and SYK model too). The first attempt of progressive time average we are going to describe is given by the following formula
\begin{align}\label{eq:PTA_naive_definition}
&\bar g (t; \,\beta) \equiv \frac 1t \int_0^t \, g(t^\prime ; \, \beta) \, d t^\prime \ ,
\end{align}
in which the average is obtained by integrating the spectral form factor from $0$ to $t$ and then dividing by $t$. We plot the progressive time average of the SFF for the GW model, together with the SFF itself, in Figure \ref{fig:naive_PTA_GW_beta01}. 
 We see that the progressive time average reaches for late times a plateau which reproduces the behavior of the SFF. However for early times it cannot reproduce the ``slope-ramp'' behavior (for $\beta = 0.1 , \, 1$). We observed an identical behavior for SYK and hence we conclude that the failure of the progressive time average (\ref{eq:PTA_naive_definition}) to reproduce the early time behavior of the spectral form factor is a general feature which goes beyond to the particular case of GW model.

\begin{figure}
\centering
    \includegraphics[width=0.6\textwidth]{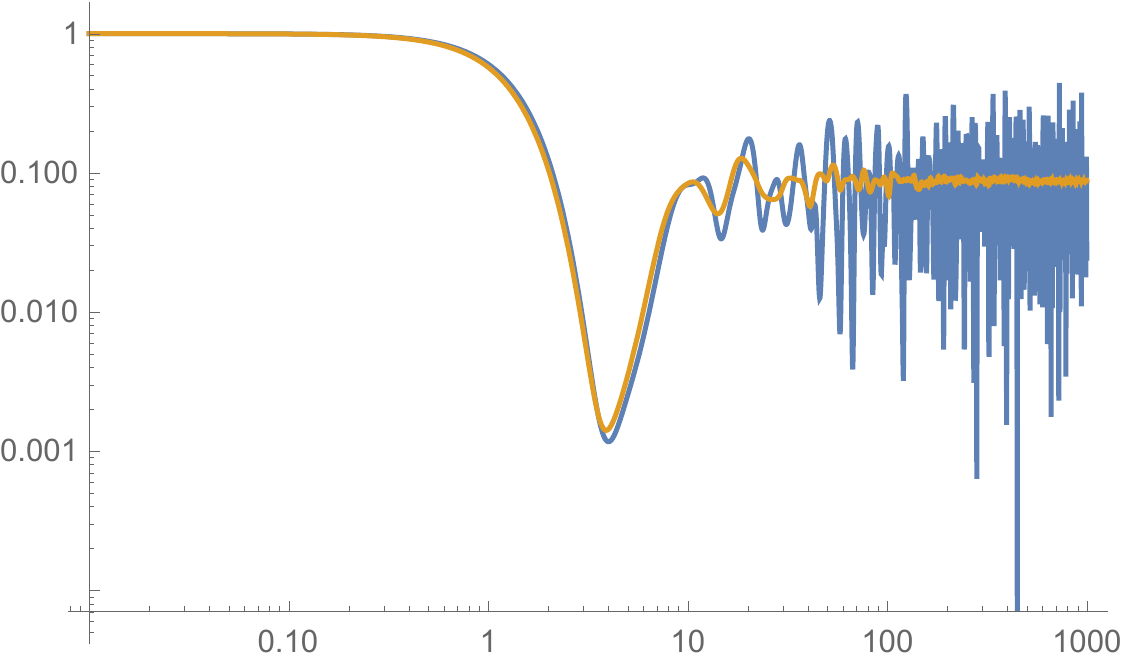}
   \caption{The progressive time average (\ref{eq:PTA_improved_definition}) for the GW model with $N=32$. $\beta = 0.1$}
    \label{fig:improved_PTA_GW_beta01}
\end{figure}

What we found is in agreement with the heuristic arguments of \cite{Balasubramanian:2016ids}: whereas the late time behavior of the progressive time average is pretty independent on the details of the time window, the early time behavior strongly depends on these details. Hence, we should adjust the parameters of the time window in order to get a better agreement between the SFF and its progressive time average. After some trial and error we found that the progressive time average
\begin{align}\label{eq:PTA_improved_definition}
&\bar g (t; \,\beta) \equiv \frac {1}{0.3 t} \int_{0.9 t}^{1.2 t} \, g(t^\prime ; \, \beta) \, d t^\prime \ ,
\end{align}
gives a good average of the SFF, and especially for small values of $\beta$ it is able to wash out the erratic late time behavior while preserving the initial self averaging phase, as we show in  Figures \ref{fig:improved_PTA_GW_beta01}  for the GW model with $\beta = 0.1$. 

These results give further indications that the proposal of \cite{Balasubramanian:2016ids} is very efficient especially when the single realization of the SFF is already sufficiently uniform and that, at the moment, it is the best option to study the SFF of standard QFTs for which an ensemble average is not available.

\subsection{The Spectra and the differences in the SFFs}\label{sec:spectra}

Let us summarize what we found so far: the SFF for the GW model, for low values of $\beta$, shows a ``slope-ramp-plateau" structure that resembles the analogous features we see for SYK. Apart from this qualitative similarity the plots for the two theories look different: the dip time in SYK occurs around $t \sim 5 \cdot 10^2$, whereas in the GW model the dip time is around $t \sim 3$; moreover the ramp in SYK is much longer and with a much smaller slope, whereas in the GW model the ramp looks more like a ``bump'' and the plot rebounds to the plateau after a very short amount of time. The time in which the plateau starts is around $t \sim 10$ for the GW  model and around $t \sim 10^5$ for SYK. Finally, the height of the plateau is around $10^{-5}$ for SYK and $0.1$ for the GW model. This comparison cannot be done for higher values of $\beta$: at $\beta = 10$ most of the interesting features of the plot in the GW model are already washed away.

In this Section we will provide an explanation to these differences in terms of the spectra of the two models. Once again, the main features of the spectra for both SYK and the GW model have been already discussed in previous works \cite{KSS}, \cite{You:2016ldz} and \cite{Garcia-Garcia:2016mno} and, for the case of SYK, a precise connection with RMT spectra have been worked out in \cite{You:2016ldz} and \cite{Garcia-Garcia:2016mno}. However a discussion of how the differences in the SFFs are related to the respective spectra has not been done yet, and we want to fill this gap in this Section. \footnote{We thank F.~Ferrari and M.~Hanada for interesting discussions on this point.}

\begin{figure}
\centering
    \subfigure{\includegraphics[width=0.6\textwidth]{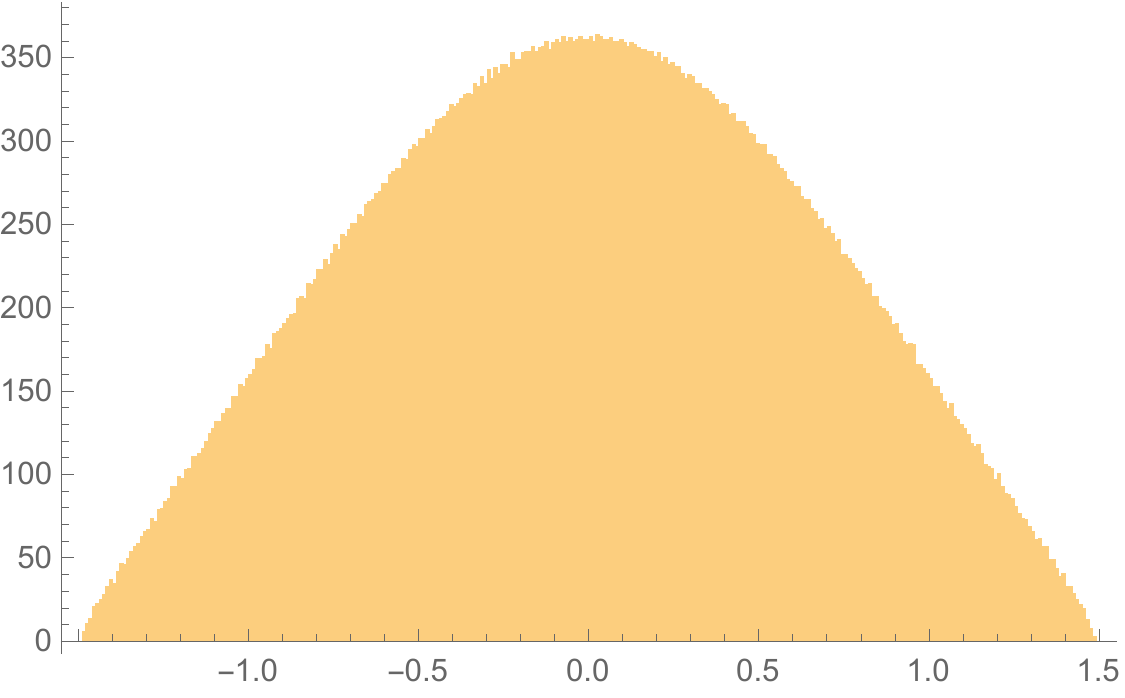}}
    \hspace{.4cm}
    \subfigure{\includegraphics[width=0.6\textwidth]{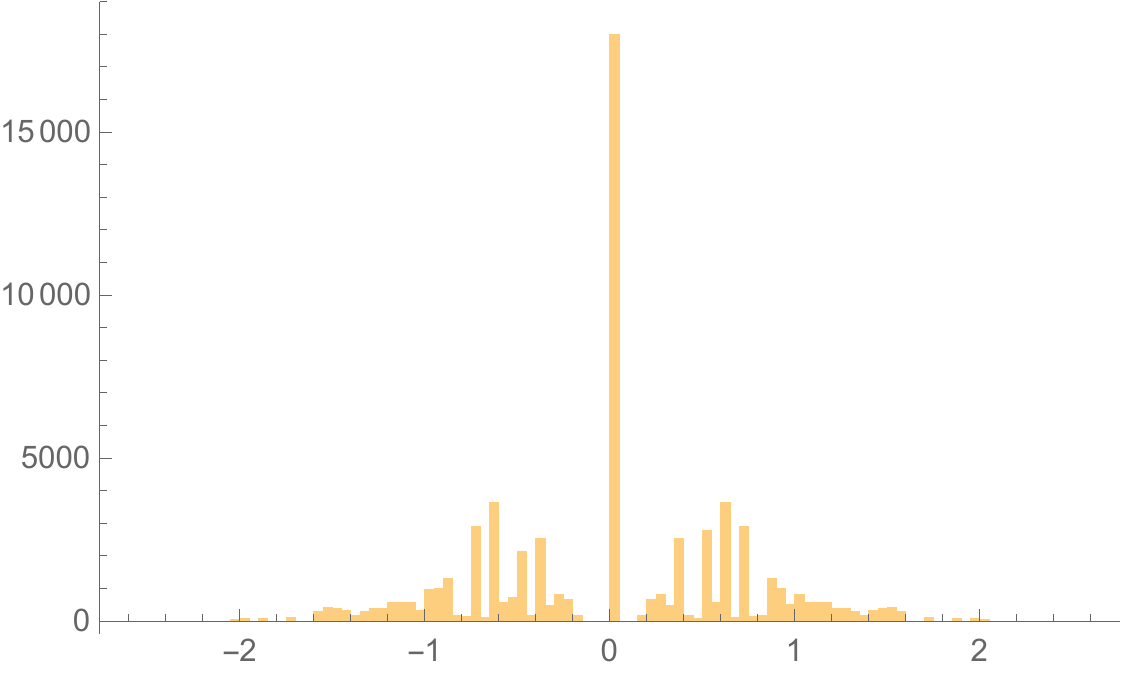}}
   \caption{Top: The SYK spectrum. Bottom: the GW model spectrum}
    \label{fig:SYK_spectrum_full}
\end{figure}

The spectrum of SYK with $N=32$ fermions  is reported in Figure \ref{fig:SYK_spectrum_full} on top.
 It has been already observed in \cite{Garcia-Garcia:2016mno} that the bulk of the spectrum, around $E=0$, follows quite closely a Gaussian distribution, while the tails of the spectrum, for finite N, have departures from the Gaussian distribution and are quite well approximated by the famous ``semicircle'' law.

The spectrum of the GW model is represented in Figure \ref{fig:SYK_spectrum_full} on the bottom. It looks completely different from the corresponding spectrum of SYK: indeed it does not have a shape similar to the semicircle law of RMTs\footnote{We have noticed however that if one removes the degeneracies, which is something one might do in order to loosely mimic the gauged theory, then the density-of-states plot has a crudely semi-circular-like form. But since the numerosity of $n=2$ model is too low, we will not comment about this further.}, rather it is characterized by many degeneracies at specific values of the energy (particularly close to the point $ E = 0$); since the numerosity is low for the $n=2, D=3$ case, we cannot find a specific trend. We observe that the ground state is not degenerate, but the first excited levels are already degenerate and they are well separated from the ground state (a feature already noticed in \cite{KSS}). It would be interesting to explain these degeneracies using the results of \cite{finite-N}, suitably generalized to the GW model. 

To understand how these differences are at the origin of the differences in the SFFs, we have to massage a little the expression for the progressive time averaged SFF. \footnote{We will consider for simplicity the ``na\" ive" expression for the SFF (\ref{eq:PTA_naive_definition}) since for our purposes it will be sufficient.} By combining the expressions (\ref{eq:SFFdefinition}) and (\ref{eq:PTA_naive_definition}) we get
\begin{align}\label{eq:massaging_PTA_one}
& \frac 1T \, \int_0^T dt \, \lvert  \frac{Z(\beta ; \, t)}{Z(\beta; \, 0)}\rvert^2 = \frac{1}{Z(\beta; \, 0)^2} \, \sum_{m , n} N_{E_m} N_{E_n} \, e^{- \beta (E_m + E_n)} \, \frac 1T \int_0^T dt \, e^{i(E_m - E_n) t} \ ,
\end{align}
where the summation runs over all energy levels and we denoted by $N_{E_m}$ the degeneracy of the level with energy $E_m$. The integral in (\ref{eq:massaging_PTA_one}) can be performed and we finally arrive at the following, manifestly real, expression
\begin{align}\label{eq:massaging_PTA_two}
\frac{1}{Z(\beta)^2} \, \sum_m \sum_{\Delta E \geq 0} \, N_{E_m} N_{E_m + \Delta E} \, e^{- 2 \beta \, E_m} e^{- \beta \, \Delta E} \, \frac{\sin (\Delta E \, T)}{\Delta E \, T} \ ,
\end{align}
where one of the two sums runs over the energy levels, whereas the other runs over the energy gaps (taken with the positive sign). 

We can see that (\ref{eq:massaging_PTA_two}) contains the dimensionless quantities $\beta \Delta E$. These quantities, together with $\beta E_m$, control how many energy levels effectively contribute to the SFF: when they are large the contribution of the corresponding energy level to the SFF is highly suppressed. In the extreme case in which all the products $\beta \Delta E$ are large, only the ground state effectively contributes to the SFF: in this case the averaged SFF tends to a constant with some small oscillations at early times due to the (very small) effects of the oscillatory term $\sin (\Delta E \, T)$, which are rapidly suppressed by the denominator.

Another piece of information that we need to recall is the following: when the quantities $\beta \Delta E$ and  $\beta  E_m$ are small enough (or in the extreme case $\beta = 0$), such that a large portion of the spectrum contributes to the SFF, the slope of the SFF is mostly controlled by the tail of the spectrum, i.e. by the region in the spectrum closest to the ground state; on the other hand, the ramp and the plateau are mostly controlled by the bulk region of the spectrum, which is the central region \cite{shenkertalk}. Moreover, it has been explained in \cite{Cotler} that the time averaged SFF has an initial decay for early times, as we can see from the term $1/( \Delta E \, T)$. This decay continues until the spectrum can be approximated by a continuous distribution and it stops when the discrete spectrum effects become relevant. From this point of view, the dip time $T_d$ is a crossover, from the time region in which the continuous distribution approximation is valid to the time region in which the discrete spectrum effects cannot be neglected.

It is not easy to characterize the precise time in which the continuous approximation cannot be applied anymore however, from (\ref{eq:massaging_PTA_two}), one can imagine that it is controlled by the dimensionless quantity $\Delta E \, T$, computed {\it in the tail of the spectrum}. Hence, we conclude that we should have a relation like this
\begin{align}\label{eq:diptimeconj}
 T_{d \, SYK} / T_{d \, GW} \sim \Delta E_{GW} / \Delta E_{SYK} \ ,
\end{align}
where $\Delta E_{SYK}$ and $\Delta E_{GW}$ are the mean level spacings, computed for the tail of the spectra, for SYK and for the GW model, respectively. We remind the reader that $T_d$ stands for the dip time of the corresponding theories here.

We have computed the mean level spacings for the tails of the spectra plotted in Figure \ref{fig:SYK_spectrum_full} and we found:
\begin{align}
  \Delta E_{GW} / \Delta E_{SYK} \sim 2 \cdot 10^2 \ , 
\end{align}
from which, using (\ref{eq:diptimeconj}), we conclude
\begin{align}
T_{d \, SYK} / T_{d \, GW} \sim  2 \cdot 10^2 \ ,
\end{align} 
a result that agrees with the plots reported in Figure \ref{fig:comparison_SYK_GW_beta01}. 

Given the mean level spacings in the tail of the spectra, we understand why, for $\beta = 10$, the SFF for the GW model is completely oscillatory. Indeed we have
\begin{align}
\Delta E_{GW} \sim 0.05 \ ,
\end{align}
and so we see that, for $\beta = 10$, the quantity $\beta \, \Delta E_{GW} \sim 0.5 $ is already big enough such that just few energy levels can be probed.

It is possible also to give an estimate of the plateau time. Indeed the plateau appears when the quantities $\Delta E \, T$, computed {\it in the bulk of the spectrum}, become big. Indeed, when these quantities are big the corresponding term gets suppressed and the progressive time averaged SFF tends to 
\begin{align}
\frac{1}{Z(\beta)^2} \, \sum_m  N_{E_m}^2 \, e^{- 2 \beta \, E_m} \ ,
\end{align}
this constant value is the height of the plateau. Hence we get the following relation
\begin{align}
T_{\mathrm plateau} \sim 1 / \Delta E \ ,
\end{align}
with $\Delta E$ being the mean energy gaps computed in the bulk of the spectra. We have computed the bulk mean energy gaps and we found
\begin{align}
\Delta E_{SYK} \sim 1 \cdot 10^{-5} \ , \qquad \Delta E_{GW} \sim 0,03 \ ,
\end{align}
from which we deduce
\begin{align}
T_{\mathrm{plateau}}^{SYK} \sim 10^5 \ , \qquad T_{\mathrm{plateau}}^{GW} \sim 30 \ ,
\end{align}
and we see that both the results are in good agreement with the plots of the SFF in Figure \ref{fig:comparison_SYK_GW_beta01}. 

Even if the agreement is quite good, the reader should note that our estimate would provide a plateau time which is {\it independent} on $\beta$. This is not correct: as explained in \cite{Cotler}[App.~H] the plateau time has a dependence on $\beta$ and this dependence is more evident by passing from $\beta = 1$ to $\beta = 5$. We did not find such a dependence in our discussion since we considered the {\it average} of the energy gaps in the bulk. Such an approximation is good enough for SYK (and for low values of $\beta$) but we will see that it is not accurate for the generalized SYK model that we will study later.    

From the results just described we understand that the reason why the ramp of the SFF for the GW model is not very long (and it looks like a rebound) is due to the fact that the energy gaps $\Delta E$ in the tail and in the bulk of the GW spectrum are very similar. Hence, the time in which the continuous approximation stops its validity is very similar to the plateau time. More generally, we understand that the length of the ramp must be related with the ratio $\Delta E_{\mathrm{tail}} / \Delta E_{\mathrm{bulk}}$: if the average energy gaps, computed in the tail and in the bulk, are very different the ramp will be longer, if they are comparable the ramp will be shorter. By global rescalings of the hamiltonian, one can change the dip time and plateau time, but the ratio $\Delta E_{\mathrm{tail}} / \Delta E_{\mathrm{bulk}}$ is invariant under rescalings: this means that the length of the ramp is not affected by global rescalings of the hamiltonian. 

We stress that, as noticed in \cite{Cotler}, the ramp is due to {\it non-perturbative} effects in these models: hence, we deduce that this hierarchy between the mean energy gaps in the tail and in the bulk of the spectrum must be controlled by some non-perturbative phenomenon. It would be extremely interesting to investigate this point further in the future.

One final comment we will make in this subsection is that various features of the SFF (in particular the existence of a clear plateau) is related to the existence of the degeneracies in the $n=2, D=3$ GW model that we consider here\footnote{We are not suggesting that the degeneracies are the only way a plateau can show up: of course, in SYK in many cases there are no degeneracies. But we have noticed that removing the degeneracies in this particular model removes the well-defined plateau.}. For holographic purposes, it will certainly be more interesting to study the gauged version of the theory, where we expect the degeneracies to be absent\footnote{We are talking about the even $n$ theory here, the odd $n$ theory is anomalous. Note also that some of the energy levels in the ungauged theory might not be present in the gauged theory, because of the absence of singlets in those levels.}. In such a case, it will almost certainly be necessary to go to higher $n$ in order to study these features reliably, to have enough numerosity in the spectrum. We will report on some related work in the KT model elsewhere.

\section{A preliminary analysis of the generalized SYK model}
\label{sec:coloredSYK}

In this Section we perform a preliminary analysis of the RMT properties of the generalized SYK model, as introduced by Gross and Rosenhaus in \cite{Gross:2016kjj}. The generalized SYK model should be thought as a family of models, parametrized by 3 sets of numbers. We will start by recalling the main features of this model and then we will focus our attention on a particular set of parameters. The choice of the parameters will be dictated by the fact that, as we will see, for this set of parameters the spectral properties of the model will be quite different from SYK. Interestingly, some contact points with the GW model have been already noticed in the literature \cite{Bonzom:2017pqs}, \cite{KlebanovCount} and in this Section we will show the RMT counterpart of that observations: the generalized model and the GW model follow the same RMT pattern. 

\subsection{A brief review of the model}
\label{subsec:review}

The generalized SYK model is constructed as follows. Let us introduce a flavor index $a$, with $a= 1 , \dots , f$. Let us also say that we have $N_a$ fermions for each flavor, such that we have $N = \sum_{a=1}^f N_a$ fermions in total, and that each flavor appears $q_a$ times in the $q$-terms interaction, $q = \sum_{a=1}^f q_a$. The action for the generalized model is
\begin{align}
\label{eq:fSYK_action}
S_{\mathrm{f SYK}} = \int \, d \tau \left(\frac 12 \sum_{a= 1}^f\,\sum_{i=1}^{N_a} \, \chi^i_a \partial_t \chi^i_a + \frac{(i)^{\frac q2}}{\prod_{a=1}^f q_a!} \sum_{I}^N J_{I} (\chi^{i_1}_1 \cdots \chi^{i_{q_1}}_1) \cdots (\chi^{j_1}_f \cdots \chi^{j_{q_f}}_f) \right) \ ,
\end{align}
where $I$ is a collective index, $I \equiv i_1, \dots , i_{q_1}, \dots , j_1 , \dots , j_{q_f}$. The random couplings $J_{I}$ are extracted from a Gaussian distribution with vanishing mean value, variance
\begin{align}
\langle J_I J_I \rangle = J^2 \, \frac{\sum_{a=1}^f N_a}{\prod_a N_a^{q_a}} \prod_a (q_a -1) ! \ ,
\end{align}
and they are anti-symmetric under permutations of indices in the same flavors. We also define
\begin{align}
\label{eq:notations_fSYK}
 \kappa_k \equiv \frac{N_k}{N} \ , \qquad Q_k \equiv \prod_{a \neq k} q_a \ .
\end{align}
The generalized SYK models (\ref{eq:fSYK_action}) are then specified by $3$ sets of numbers: the number of flavors $f$, the integers $q_k$ and the set of continuous numbers $0<\kappa_k<1$. The standard SYK models are characterized by $f = 1$, $q_1 = q$, $\kappa_1 = 1$.

After the disorder average, the standard SYK model is characterised by an $O(N)$ symmetry. In the generalized models, this symmetry gets reduced to $O(N_1) \times O(N_2) \times \cdots \times O(N_f)$.

We will perform some preliminary spectral analysis for a particular example of the generalized SYK model: we will consider the case with $f= 4$ , $q_i = 1$ and $N_i = N / 4$. and, once again, we will set $J=1$.  For this particular choice we get that the coupling $J_{ijkl}$ does not have any particular symmetry property and its variance simplifies to
\begin{align}
\langle J_{ijkl} J_{ijkl} \rangle = \frac{4^4}{N^3} \ ,
\end{align}
whereas the Hamiltonian is
\begin{align}
\label{eq:generalizedhamiltonian}
H = \sum_{ijkl = 1}^{N/4} J_{ijkl} \, \chi_1^i \chi_2^j \chi_3^k \chi_4^l \ .
\end{align}

We represent the fermions via Gamma matrices. We need $N$ Gamma matrices $\Gamma_i$ with $i = 1 \, \dots , N$. The fermions belonging to the first flavor $\chi_1^1 \dots \chi_1^{N_1}$ are represented via the first $N_1$ Gamma matrices $\Gamma_1 \dots \Gamma_{N_1}$; similarly the fermions belonging to the second flavors are represented via the second $N_1$ Gamma matrices $\Gamma_{N_1 + 1} \dots \Gamma_{2 N_1}$ and so on.

\subsection{The spectral form factor}
\label{subsec:SFF_generalized}

\begin{figure}
\centering
    \subfigure{\includegraphics[width=0.6\textwidth]{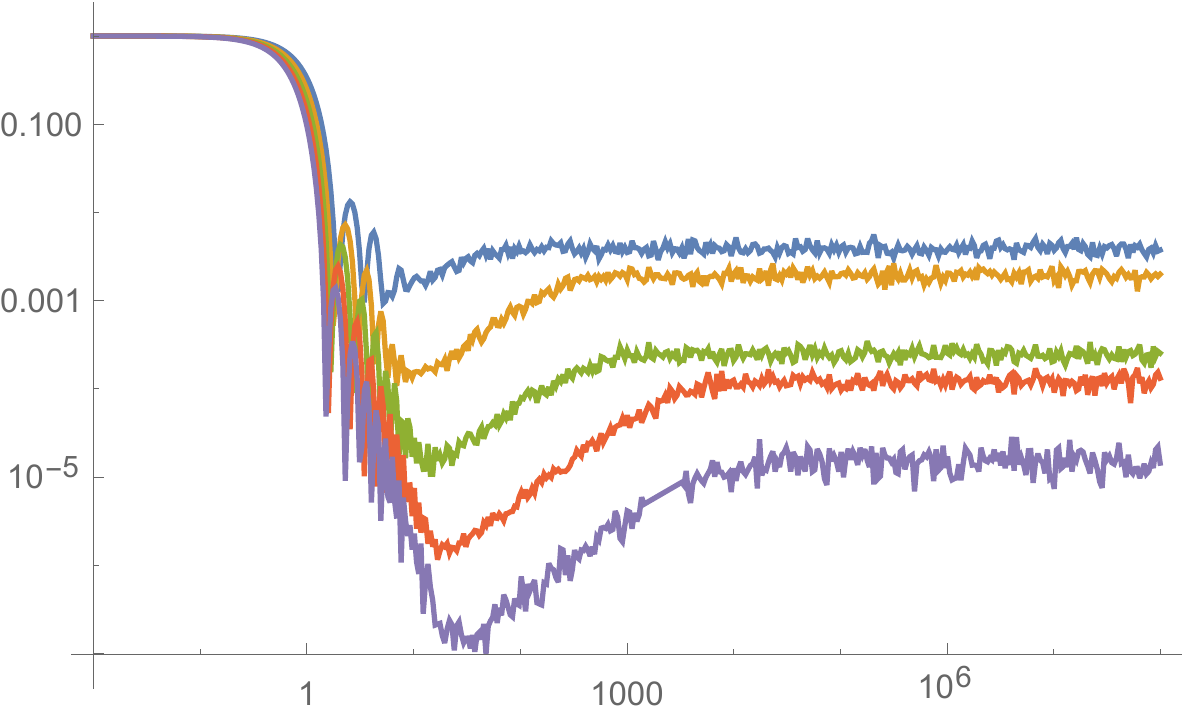}}
    \hspace{.4cm}
   \subfigure{\includegraphics[width=0.6\textwidth]{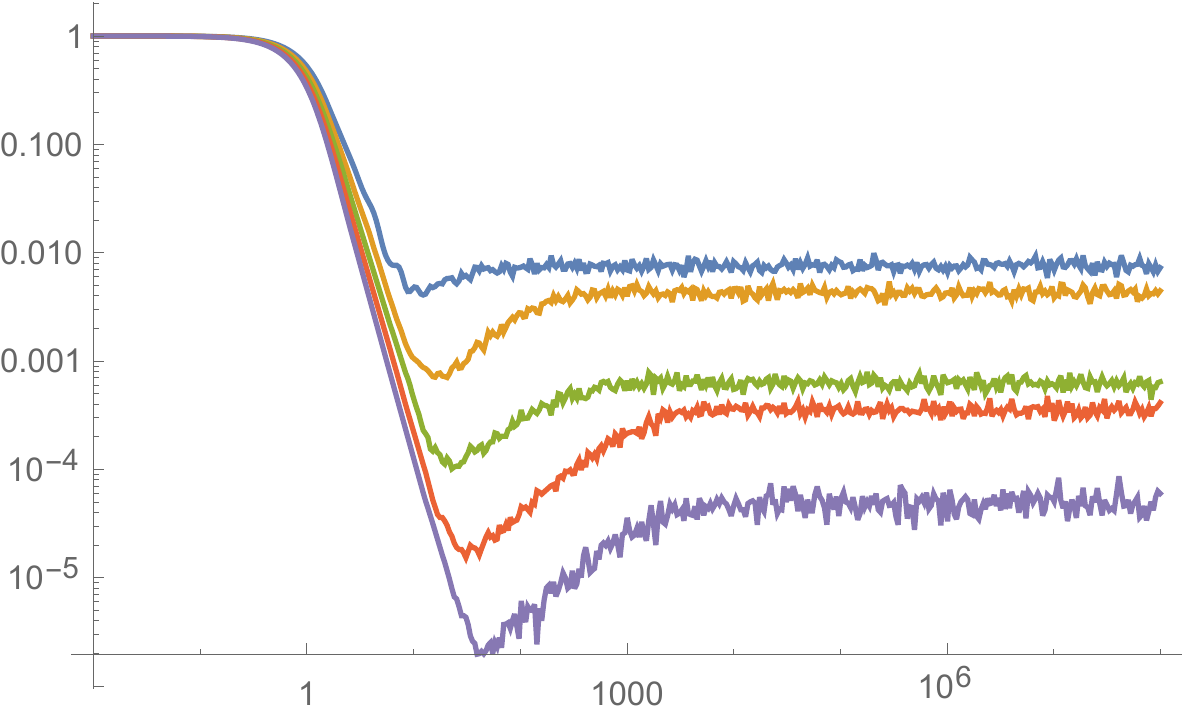}}
    \hspace{.4cm}
   \subfigure{\includegraphics[width=0.6\textwidth]{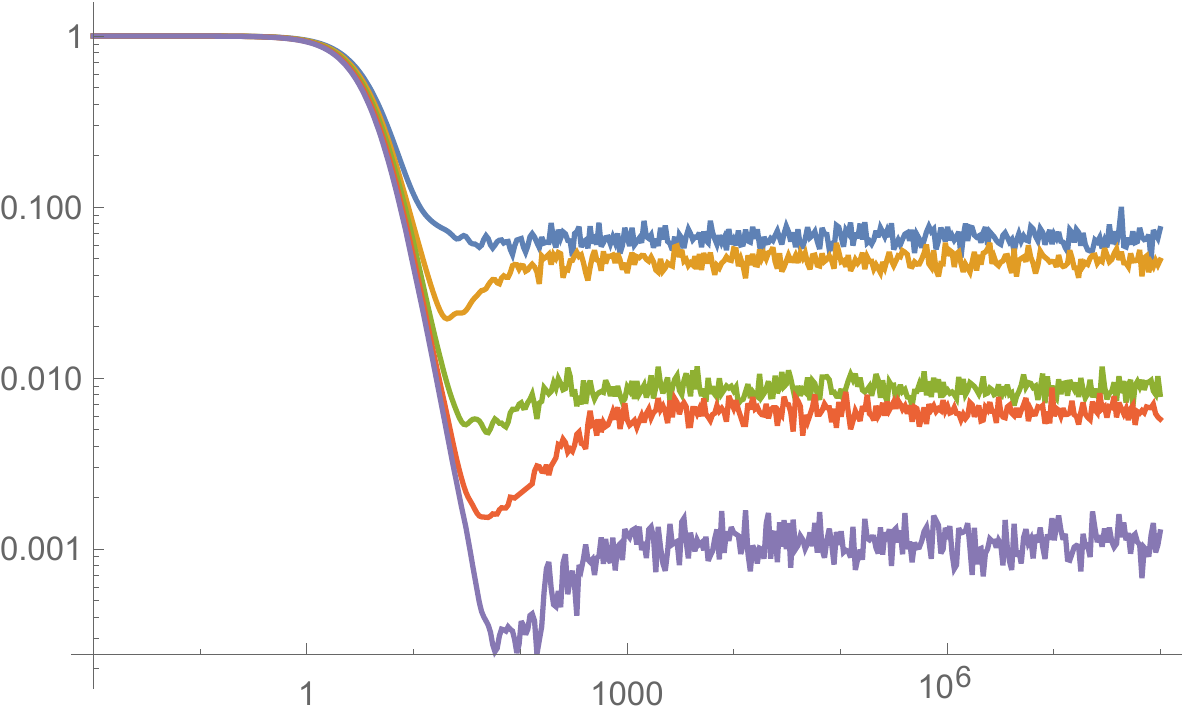}}
   \caption{The SFFs for the generalized SYK model; the values of $N$ are $16$ (blue line), $20$ (yellow line), $24$ (green line), $28$ (red line) and $32$ (violet line). Top: $\beta = 0$, center: $\beta = 1$, bottom: $\beta = 5$}
    \label{fig:SFFcolored_beta0}
\end{figure}

We have diagonalized numerically the generalized Hamiltonian (\ref{eq:generalizedhamiltonian}) for the particular values of $N = 16$,  $20$, $24$, $28$, $32$ and we computed the corresponding SFFs (\ref{eq:SFFdefinition}) for the inverse temperatures $\beta = 0$, $1$ and $5$. The plots for the ensemble averages of the SFFs are presented in Figure \ref{fig:SFFcolored_beta0}.
 
We see some peculiarities w.r.t the analogous SFFs studied for SYK  in \cite{Cotler}. The dip-ramp-plateau features are evident at very low values of $\beta$: when $\beta = 0$ we see that the ramp becomes more and more prominent by increasing $N$. However, for $\beta = 5$ this feature is less evident. In particular for the cases $N=16$, $24$ and $32$ the ramp disappears ($N= 16$ and $N= 24$) or it becomes much shorter. We will confirm that these values of $N$ are peculiar in Section \ref{subsec:generalizedspectrum}, even if we do not have an explanation of this phenomenon.
 
\begin{figure}
\centering
    \subfigure{\includegraphics{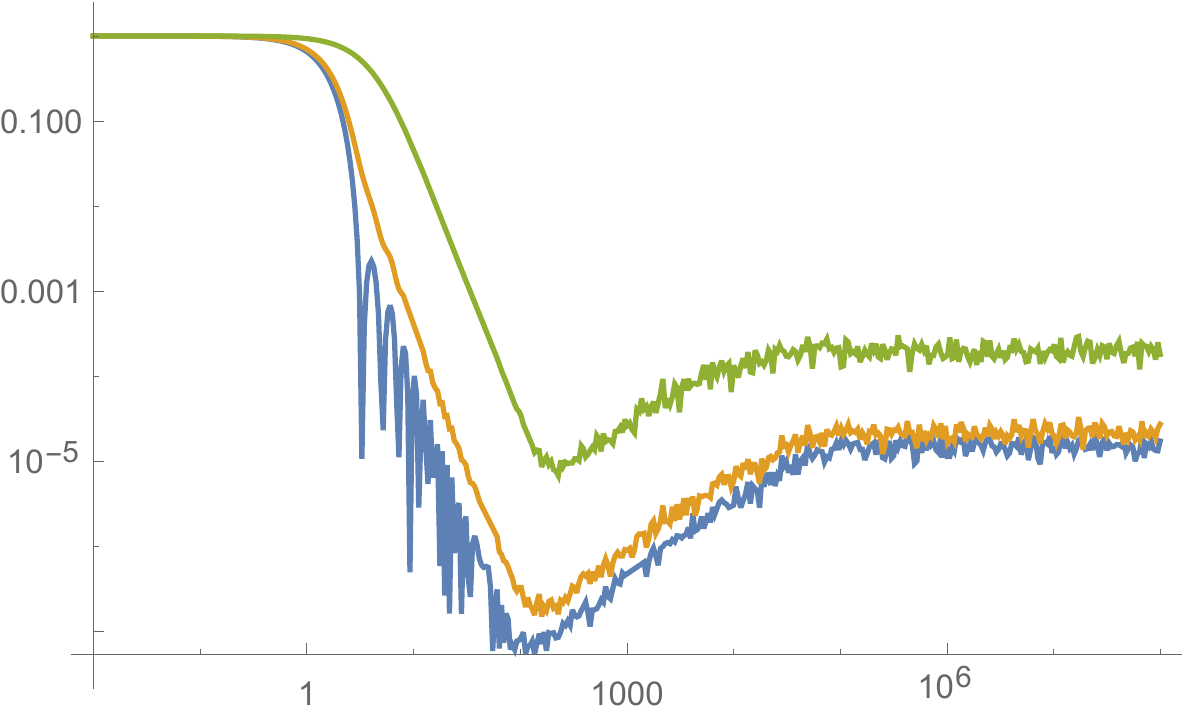}} 
   \subfigure{\includegraphics{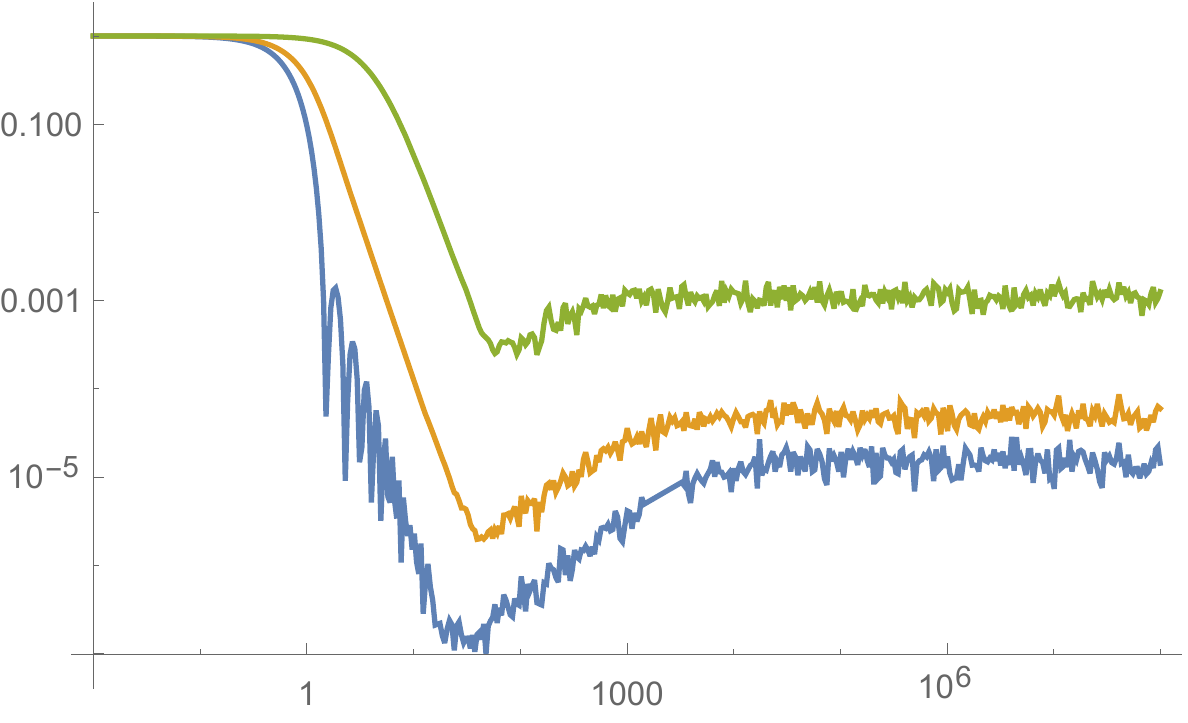}}
   \caption{Top: The SFFs for the SYK model at $\beta = 0$ (blue line), $\beta = 1$ (yellow line) and $\beta = 5$ (green line). Bottom: The SFFs for the generalized SYK model at $\beta = 0$ (blue line), $\beta = 1$ (yellow line) and $\beta = 5$ (green line).}
    \label{fig:SFFcomparison_SYK} 
\end{figure}

  In general, we notice that the ramp and the plateau time are more affected by the value of $\beta$ than in standard SYK, and for $\beta = 5$ the plateau time is much smaller than the plateau time for $\beta = 0$. Hence, we conclude that the rough analysis we performed in Section \ref{sec:spectra} to estimate the plateau time is not accurate enough for the generalized SYK model and it works well for $\beta = 0$ only.  We plot a comparison between the SFFs, for various values of $\beta$, for both the standard SYK model and for the generalized one in Figure \ref{fig:SFFcomparison_SYK}.
 It would be interesting to obtain a precise relation for the dip time and the plateau time for the generalized model, along the lines of \cite{Cotler}. We plan to address this point elsewhere. Qualitatively, we can understand why our rough estimate of the plateau time, based on the discussion of Section \ref{sec:spectra}, is valid only at $\beta = 0$: for vanishing $\beta$ we see from the formula (\ref{eq:massaging_PTA_two}) that the SFF can probe all the energy levels with the same weight $e^{- \beta E_m} = 1$. In this case, using the averaged energy gap as a quantity to estimate the plateau time gives a good estimate. When $\beta$ becomes large, the approximation using the the average energy gap becomes worst and worst, because the weights $e^{- \beta E_m}$ can vary a lot. The approximation is worst when the spectrum is more sparse, in the sense that the eigenvalues are distributed on a larger set of values. Indeed, we checked that the spectrum of the generalized model is sparser than the spectrum for the standard SYK.  Like in SYK, the dip time is almost unchanged by varying $\beta$.
  
\begin{figure}
\centering
    \subfigure{\includegraphics[width=0.6\textwidth]{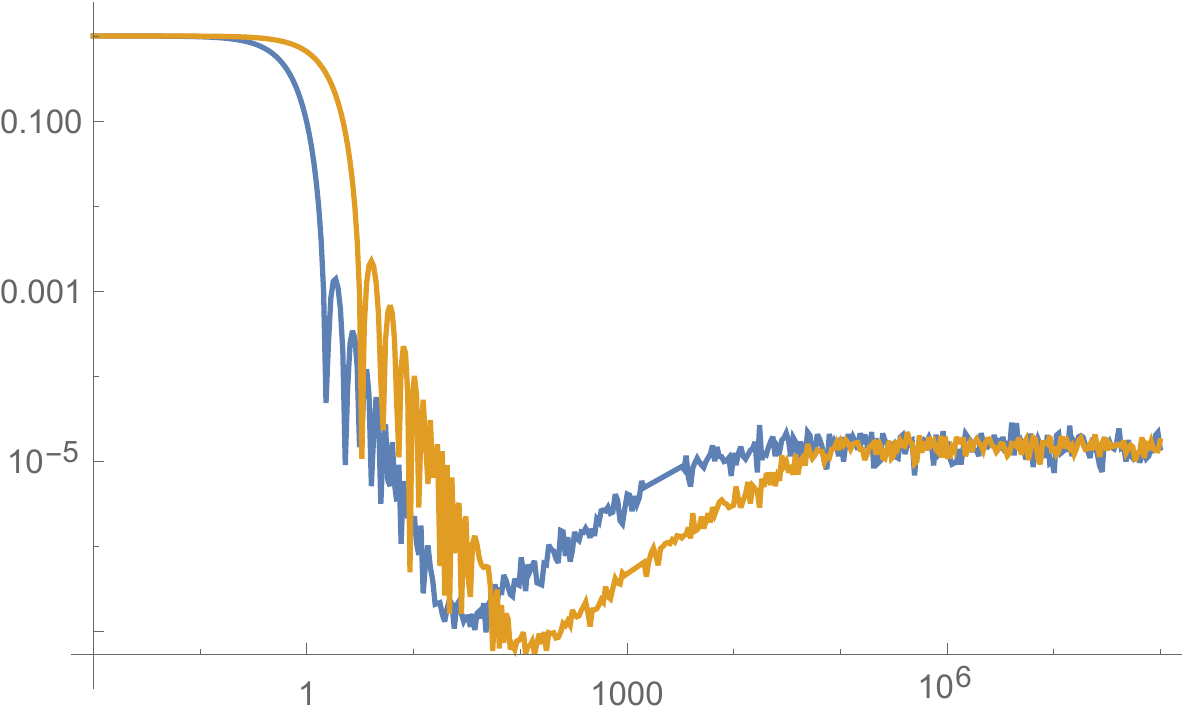}}
    \hspace{.4cm}
   \subfigure{\includegraphics[width=0.6\textwidth]{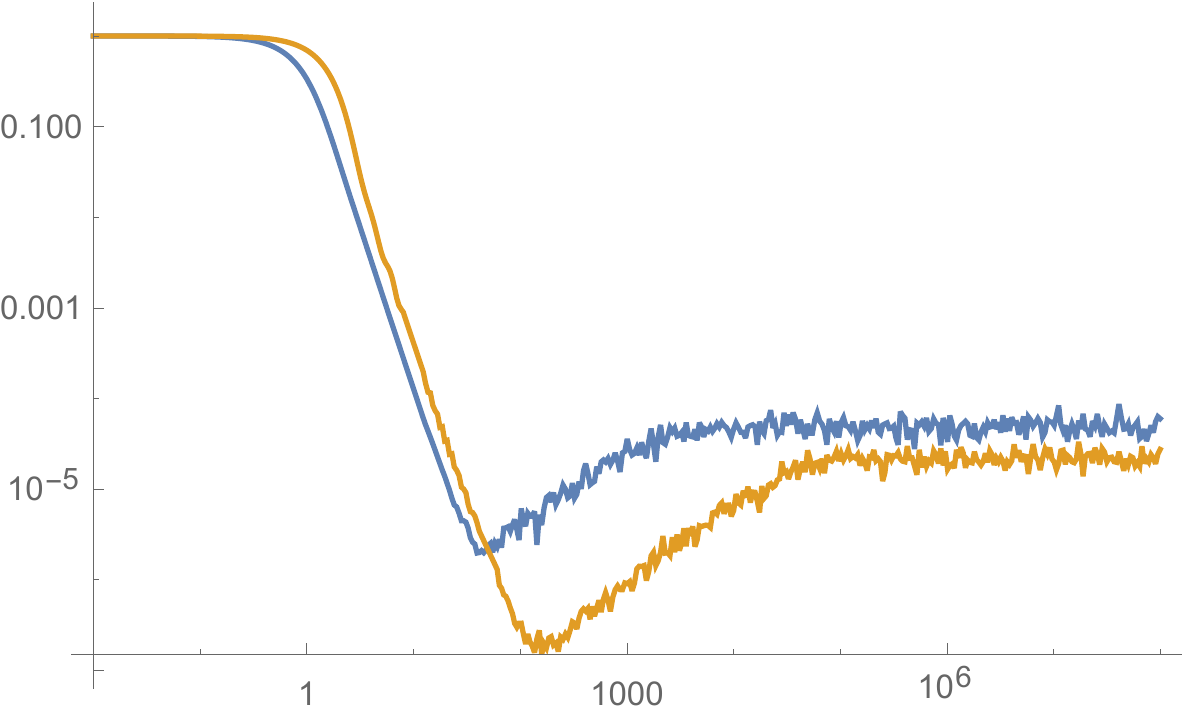}}
    \hspace{.4cm}
   \subfigure{\includegraphics[width=0.6\textwidth]{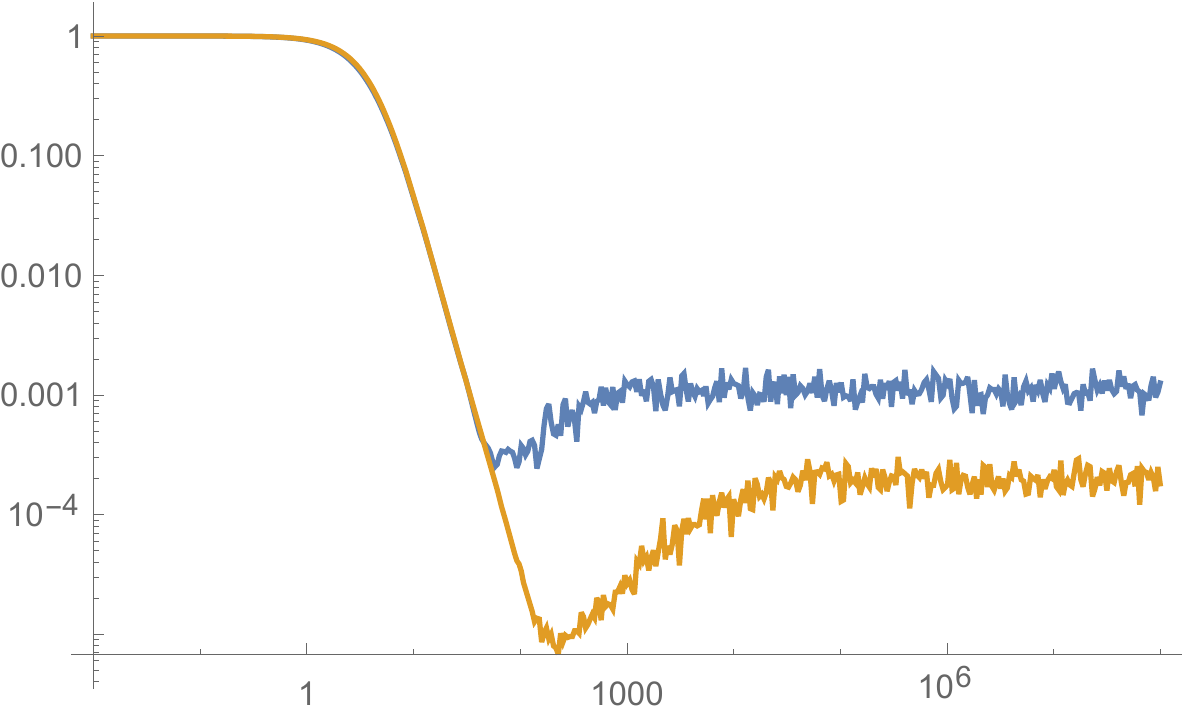}}
   \caption{The SFFs for the generalized SYK model (blue line) and for the standard SYK model (yellow line). Top: $\beta = 0$, center: $\beta = 1$, bottom: $\beta = 5$}
    \label{fig:SFFcomparison_beta0}
\end{figure}

  It is also interesting to compare, for a given value of $N$, the SFFs for SYK and for the generalized model on the same plot. We presented this comparison in Figure \ref{fig:SFFcomparison_beta0} for the particular case $N=32$. 
Once again, we see that for $\beta = 0$ the SFF of the generalized model has a behavior very similar to the behavior of the analogous SFF for SYK. The small discrepancies in the dip and the plateau times can be explained in terms of the respective spectra and in terms of the simplified analysis of  Section \ref{sec:spectra}. This is not true anymore for larger values of $\beta$: we see that in this case the ramp become shorter and shorter and the height of the plateau becomes higher. The fact that the height of the plateau is higher is telling us that the gravity dual of the generalized model (whatever it is) has an entropy which is smaller than the gravity dual of standard SYK. It is interesting to notice also that, for $\beta = 5$ the early time behaviour of the SFF is exactly the same as in the SFF for SYK. By recalling that this part of the SFF is correctly reproduced by the Schwarzian action \cite{Cotler}, we conclude that the same low energy effective description is valid also the generalized model.

\subsection{The spectrum}
\label{subsec:generalizedspectrum}

Ley us analyze some features of the spectrum for the generalized model (\ref{eq:generalizedhamiltonian}). The first and simplest observation is that the Hamiltonian (\ref{eq:generalizedhamiltonian}) enjoys {\it spectral mirror symmetry} as the GW model. We will show this property, together with a RMT classifications of the generalized hamiltonians, in the next Section. 

The standard Gaussian RMTs (GUE, GOE and GSE) do not have spectral mirror symmetry, whereas this is a typical property of the other, non-standard, ensembles of the Altland-Zirnbauer classification. Hence we conclude that the chaotic properties of the generalized SYK model will be different in detail from the analogous properties of SYK.

\begin{figure}
\centering
    \subfigure{\includegraphics[width=0.6\textwidth]{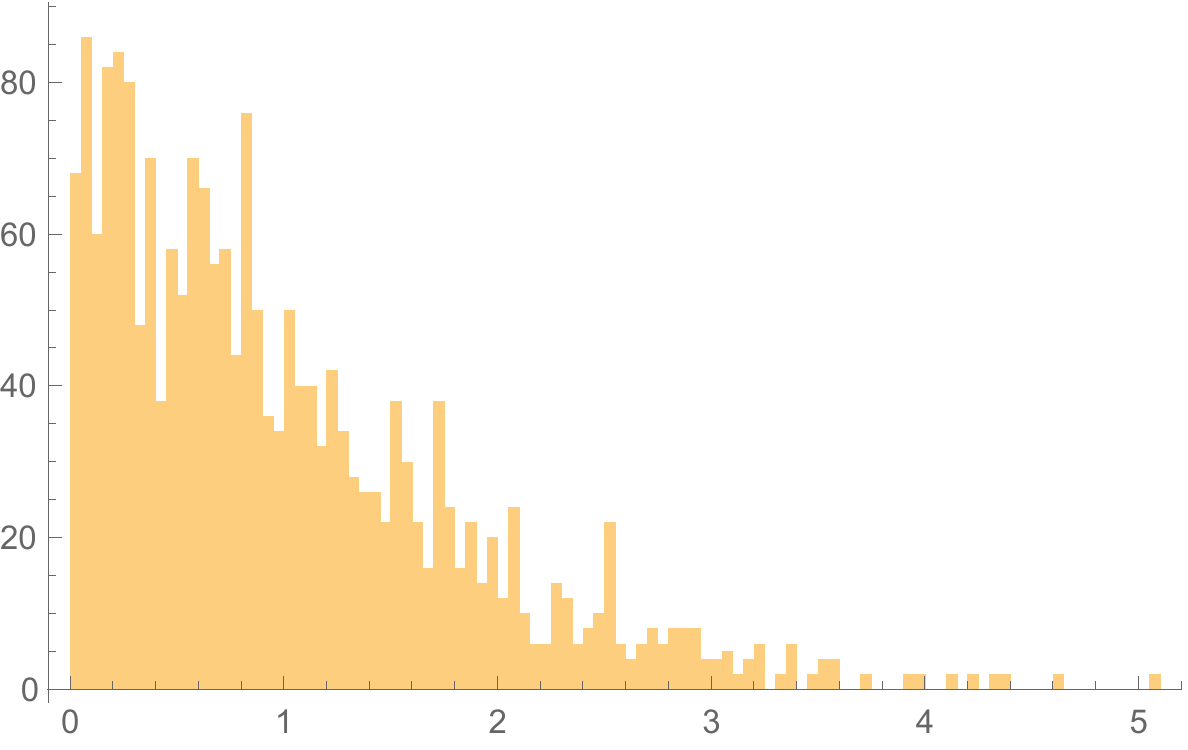}}
    \hspace{.4cm}
   \subfigure{\includegraphics[width=0.6\textwidth]{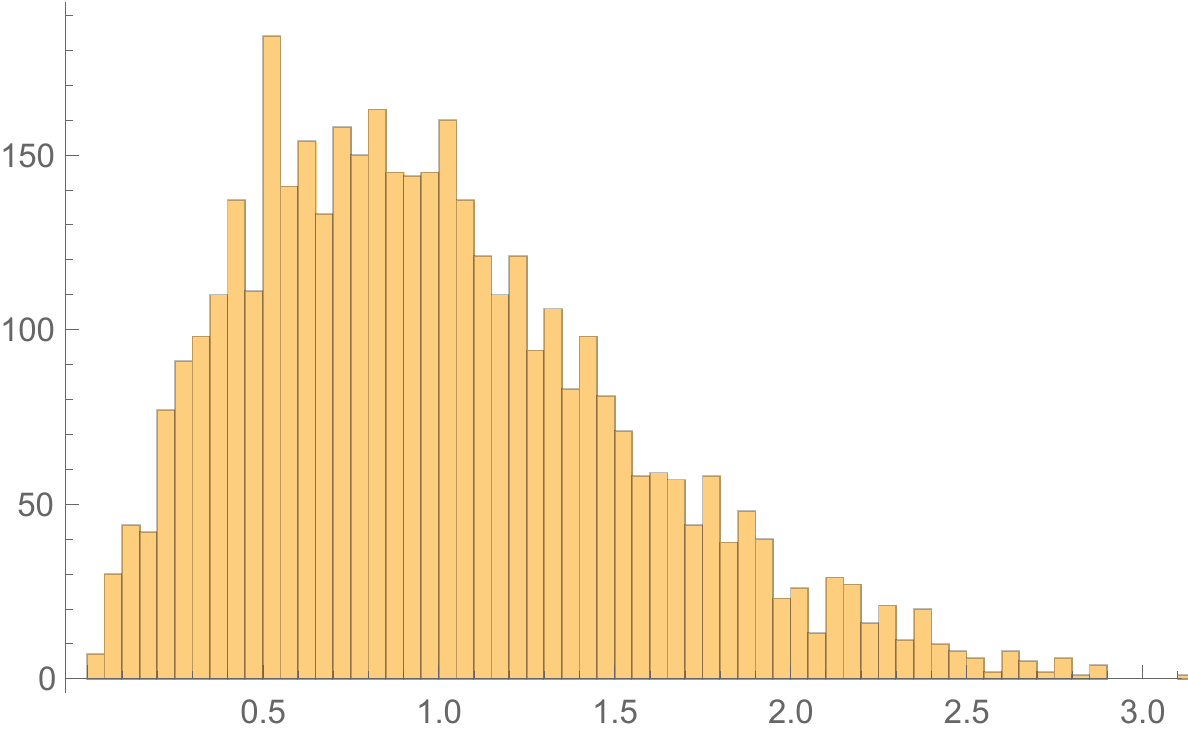}}
    \hspace{.4cm}
   \subfigure{\includegraphics[width=0.6\textwidth]{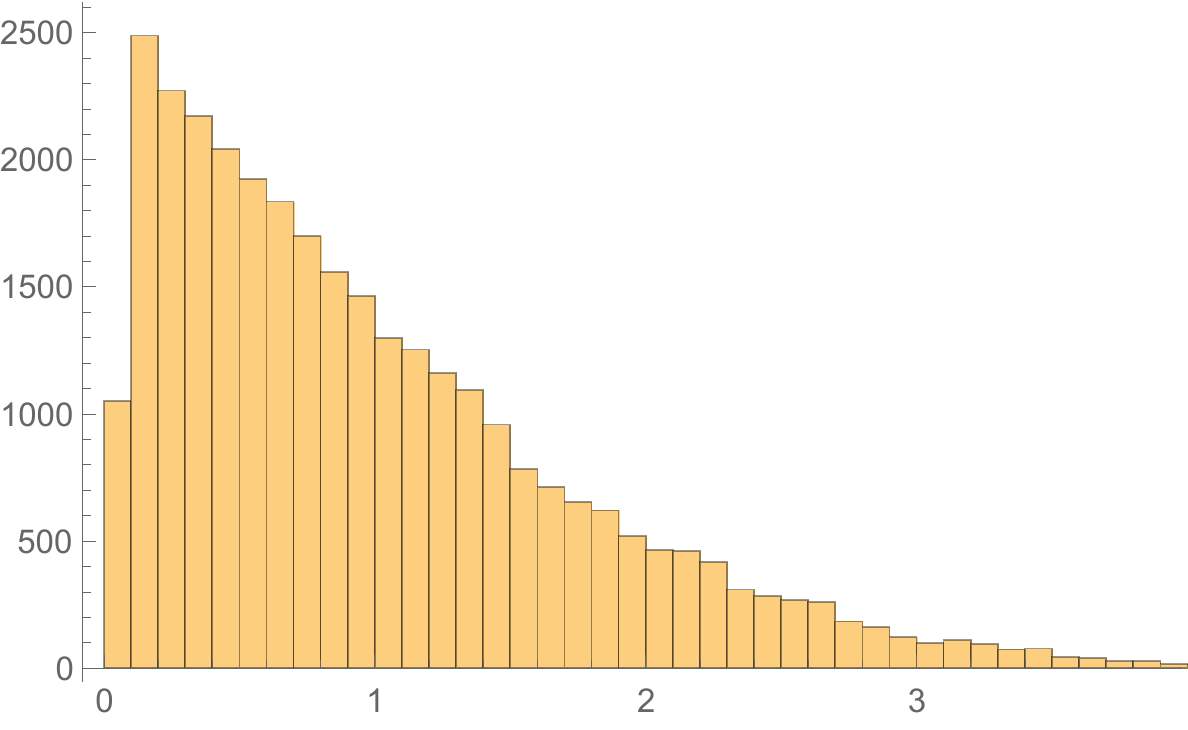}}
   \caption{The unfolded level spacing distribution for the generalized model. Top: $N = 24$, center: $N = 28$, bottom: $N = 32$}
    \label{fig:unfoldedcolored_N24}
\end{figure}

We have computed the unfolded level spacing distribution for all the values of $N$. We plot the results for the interesting cases of $N = 24$, $28$ and $32$ in Figure \ref{fig:unfoldedcolored_N24}.
For $N=24$, the plot does {\it not} show any hint of level repulsion. On the other hand the plot for $N=28$ clearly shows the repulsion, but the repulsion is still very weak for $N = 32$. From this observation, and from the plots of the SFF we already discussed, we suspect that the generalized model, for the particular case of $N \mod 8 =0$, is much less chaotic than for the other values of $N$.

It would be interesting to obtain a theoretical understanding of this phenomenon, perhaps in relationship with the properties of the non-standard RMTs. It would be also interesting to study the generalized SYK model for different values of the parameters $f$, $q_k$ and $\kappa_k$ to see if some other relevant differences with the standard SYK models can be found. We analyzed the case of $f=2$, $q_1 = q_2 = 2$ and $N_1 = N_2 = N / 2$ without finding any significant differences with SYK.

\subsection{Random matrix ensembles}
\label{subsec:RMT_ensembles}

To conclude this section, we want to understand what are the RMT ensembles to which the generalized SYK hamiltonians belong. For all the values of $N$, the generalized hamiltonian has the spectral mirror symmetry. The mirror symmetry operator $S$, satisfying the relation $\left\{ S , H \right\} = 0$, is obtained by multiplying all the fermions which belong to the first flavor. Concretely we have
\begin{align}
S \equiv \Gamma_1 \cdot \Gamma_2 \cdots \Gamma_{N_1} \ .
\end{align}
It is simple to see that $S$ is unitary $S^\dagger S = 1$ and that it anti-commutes with the hamiltonian. 

Beyond $S$, we can construct, using the same construction of \cite{KKS} and \cite{You:2016ldz}, the antiunitary operator of time reversal $\mathcal{T}$. It is explicitly realized as
\begin{align}
\mathcal{T} = P^{(N+2)/2} \, \Gamma_1 \Gamma_3 \cdots \Gamma_{N-1} \, \mathcal{K} \ ,
\end{align}
where $P$ is the chiral operator and the charge conjugation operator $\mathcal K$ satisfies
\begin{align}
\mathcal{K} \Gamma_i = - (-1)^i \, \Gamma_i \mathcal K \ .
\end{align}

The time reversal operator commutes with the generalized hamiltonian for any values of $N$. The RMT ensembles to which the generalized hamiltonians belong are determined by the squares of the operators $S$ and $\mathcal T$. By explicit computation, we find
\begin{align}
S^2 = (-1)^{\frac{N_1}{2} (N_1 -1)} \ , \qquad \mathcal T^2 = (-1)^{N_1 (2 N_1 - 1)} \ ,
\end{align}
from which we deduce the RMT ensembles to which the generalized hamiltonians belong
\begin{itemize}
\item $N_1  \mod 4 = 0$ the ensemble is the BDI class.
\item $N_1 \mod 4 = 1$ the ensemble is the CII class.
\item $N_1 \mod 4 = 2$ the ensemble is the CI class.
\item $N_1 \mod 4 = 3$ the ensemble is the DIII class.
\end{itemize}

This classification coincides with the one obtained in \cite{KKS} for the GW model. Hence, the RMT classification of the generalized SYK model is the same of the RMT classification of the GW. We believe that this observation gives the RMT counterpart of the observations made in \cite{Bonzom:2017pqs} and \cite{KlebanovCount} about the similarities between the GW model and the generalized SYK model.

\section*{Acknowledgments}

We thank F. ~Ferrari, M.~Hanada, I.~Klebanov, S.~Minwalla, E.~Witten and J.~Yoon for discussions and/or correspondence. C. K. and K. V. P thank P. N. Bala Subramanian for an early related collaboration. D.~R. is particularly grateful to J.~Kim and S.~Lee for initial collaboration on this project and for many discussions on related topics.

\appendix

\section{Glimpses of Quantum Black Holes in Riemann Zeroes} 
\label{RiemannMain}

One of the messages of SYK and related work is that the late-time behavior of quantum black holes could be controlled by random matrix/quantum chaos like behavior \cite{Cotler, KSS, KKS}. Furthermore, these models are a suggestion  that there do exist 0+1 dimensional quantum mechanical models that could capture quantum gravity in AdS$_2$. 

There exists another system which is expected to be modeled by a 0+1 dimensional quantum mechanical system that exhibits random matrix behavior and is intimately connected to quantum chaos \cite{Keating, Snaith}. This is the putative Hamiltonian whose eigenvalues are supposed to reproduce the  non-trivial zeros of the Riemann zeta function. There are two ``beliefs'' about Riemann zeros that will be relevant to our discussion in this section. One is based on the so-called Hilbert-Polya hypothesis, which says that the (imaginary part of the non-trivial) zeros of Riemann are the eigenvalues of a Hermitian Hamiltonian\footnote{An interesting proposal for a closely related Hamiltonian was recently made in \cite{Bender}. This is a bosonic Hamiltonian. SYK and related models give rise to fermionic Hamiltonians. It will be interesting to see whether some type of bosonization/fermionization type arguments can be used to relate such Hamiltonians. Note however that the Hamiltonian of \cite{Bender} is PT-symmetric rather than Hermitian, and also that its domain is not well-understood.}. The second ``belief'' is that the zeros of Riemann with very large imaginary values have quantitative features that are captured by a random unitary matrix. Both these ideas are clearly related, and form the context of our comments in this section.

In this section, we will explore this analogy and point out some connections, but also some differences. We will list out some of the features we expect from AdS$_2$ quantum gravity. Then we will discuss some of the features seen in the spectra of SYK and related tensor models. We will also compare these to some of the properties of Riemann zeors.

\begin{figure}
\centering
\includegraphics[scale=1]{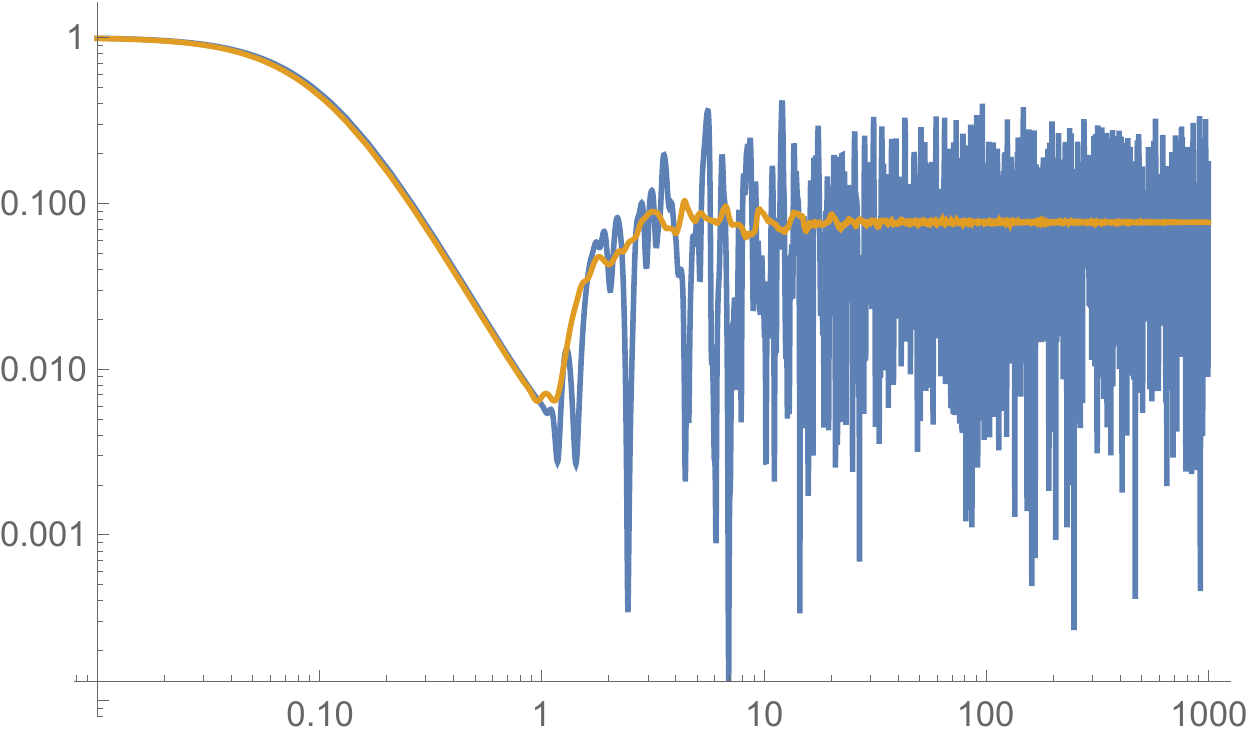}
\caption{Spectral form factor (and its progressive time average) for $\beta =0.1$ and 20000 zeros}
\label{sff10000}
\end{figure}

\begin{figure}
\centering
\includegraphics[scale=1]{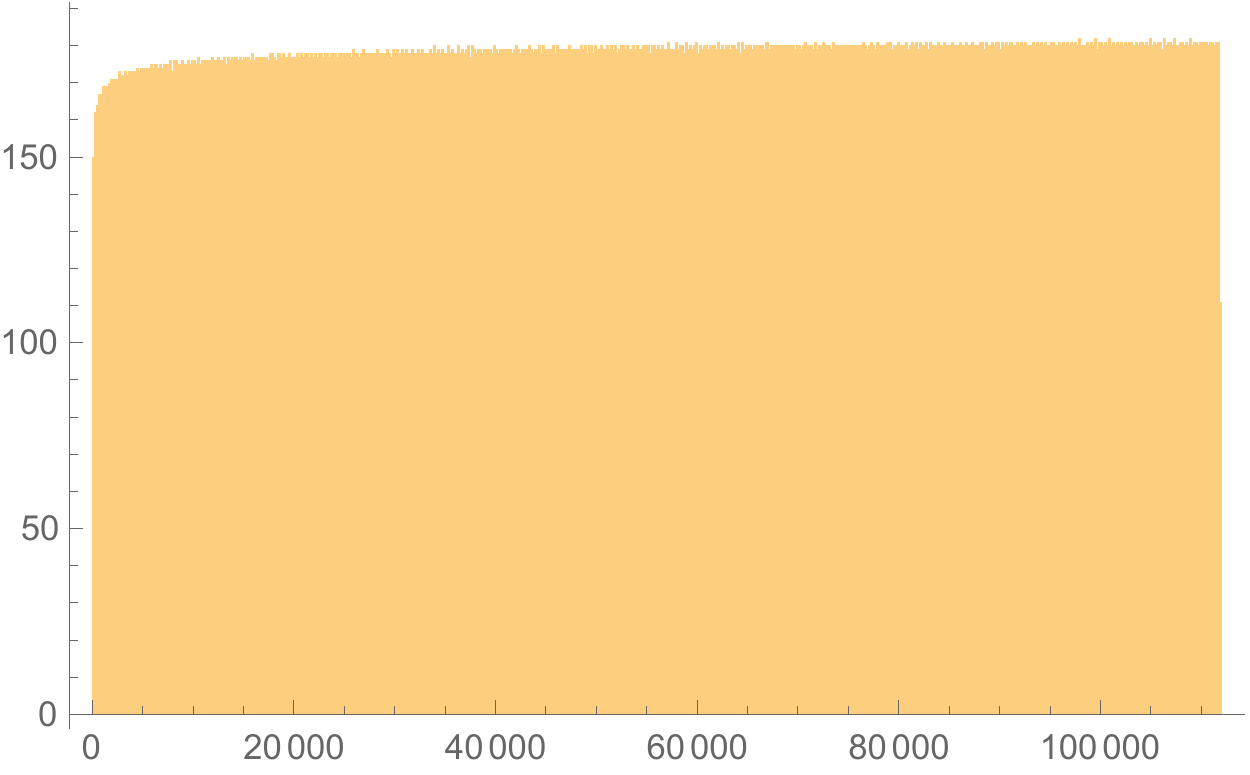}
\caption{Density of states for the first 100000 zeroes. This is very close to be a constant }
\label{DOS10000_rescaled}
\end{figure}

\begin{figure}
\centering
\includegraphics[scale=1]{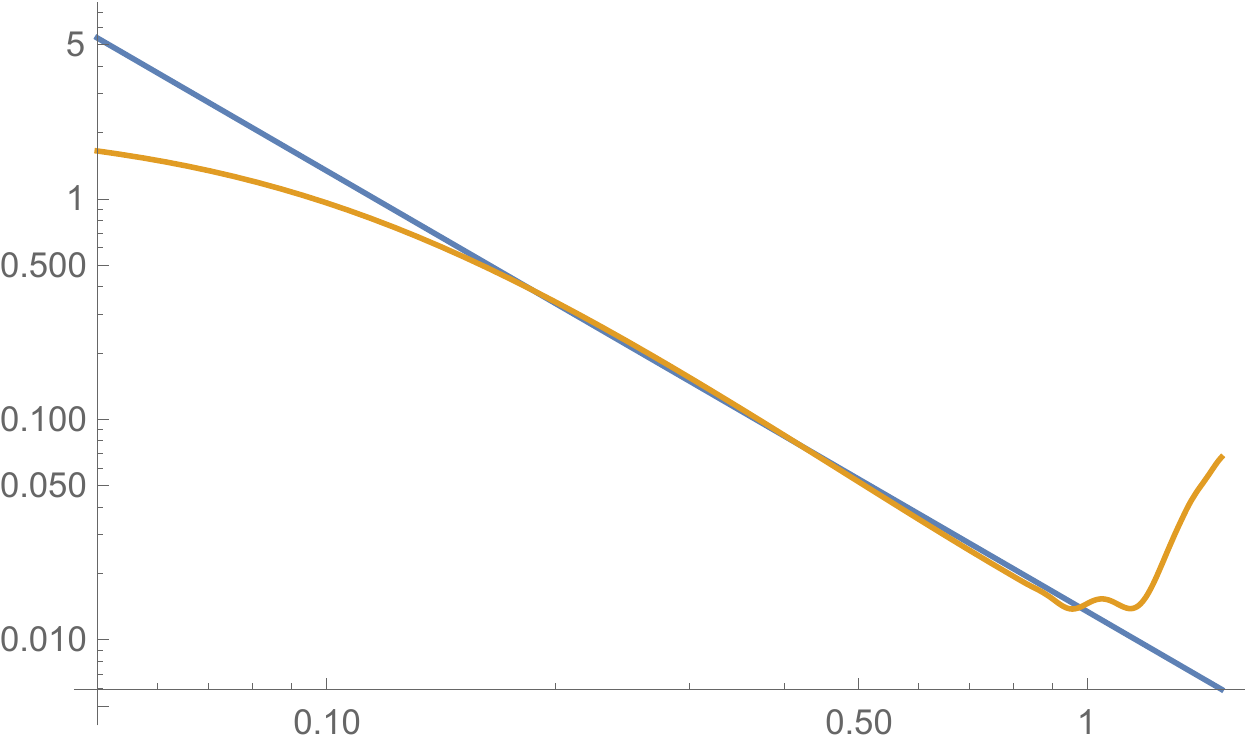}
\caption{Behaviour of the dip in the log-log plot of the SFF for the first 20000 zeroes (yellow line) plotted against the curve $y \propto \, \frac{1}{t^2}$. The agreement is very good, as expected.}
\label{dip behaviour}
\end{figure}

To start with, the Riemann zeta function 
has non-trivial zeroes at certain values $s=\frac{1}{2}+i \gamma _n$, where $\gamma _n$'s are all real according to the Riemann hypothesis. Following \cite{Snaith}, we can define the quantity
\begin{align}
w_n&=\frac{\gamma _n}{2 \pi}\log \left(\frac{|\gamma _n|}{2 \pi}\right) \ .
\end{align}
This is useful because $w_n$'s have an approximate mean separation\footnote{This accomplishes a version of unfolding of the zero spectrum. But in contra-distinction to unfolding in generic systems, which depends on the entire spectrum typically, here it is a pointwise definition which is what makes it interesting.} of unity (see figure) and the statistics of these redefined non-trivial zeroes is known to emulate the statistics of eigenvalues of random matrices \cite{Snaith}. More precisely, the pair correlations of the zeroes of Riemann zeta function and that of random unitary matrices have a similar form. The implication is that the level spacing distribution of Riemann zeroes matches the level spacing distribution (and therefore level repulsion) of GUE. These facts together imply that the late time behavior of Spectral Form Factor (SFF) of Riemann zeroes should behave like that of the SFF of random unitary matrices. The behavior of SFF was used as a diagnostic of late-time random matrix behavior in black holes in the context of SYK-like models, so this observation suggests that the spectrum of Riemann zeros also shares this property. In the plot of the SFF the dip-ramp-plateau structure is clear, even though in our numerics we are using only a fairly small number of zeros. We plot spectral form factor (and its progressive time average) for the first 20000 (rescaled) Riemann zeroes in figure \ref{sff10000}, by taking $\beta = 0.1$. 
 Note that the plateau height matches with the expectation of $\frac{Z(2\beta )}{Z(\beta )^2}$. 
The dip is dominated by the disconnected part of the SFF and can be calculated as follows:
\begin{align}
F^{(d)}_{\beta =0,t}&=\left| \int dw ~\tilde{\rho} (w) e^{-iwt}\right| ^2
\end{align}
where $\tilde{\rho} (w)$ is the mean density and is roughly a constant in the Riemann zero case as can be seen in the figure \ref{DOS10000_rescaled} for the first 100000 zeros. Hence, we see that the slope scales\footnote{Note that this behaviour is different from the $\frac{1}{t^3}$ scaling that is expected if the mean density is given by the Wigner's semicircle law. This means that the late time (but before dip) behavior of the putative Riemann Hamiltonian is {\em distinct} from what is expeected of random matrices.} like $\frac{1}{t^2}$ at late times before the dip. This can be seen in the figure \ref{dip behaviour}.

The spectral form factor at times later than dip is dominated by the connected part and can be calculated as follows:
\begin{align}
F^{(c)}_{\beta =0,t }&=\int d\lambda d \lambda ' ~R_2(\lambda ,\lambda ')e ^{i(\lambda -\lambda ')t}
\end{align} 
where $R_2(\lambda ,\lambda ')$ is the connected pair correlation function. As mentioned earlier, the pair correlation function is the same for the eigenvalues of random unitary matrices and the spectrum of Riemann zeroes. Hence, we expect that, in both the cases, the ramp and plateau have a similar behaviour, which we can see explicitly from the behavior of SFF. 

Note that the SFF for the Riemann zeros, that we plot with $20000$ eigenvalues, does not follow too closely the SFF for the GUE ensemble, especially right after the dip. The reason is that, according to some numerical evidence \cite{Snaith}, the agreement between the pair correlation of the GUE ensemble and that of the Riemann zeros is only true in the limit in which we take Riemann zeros with much larger imaginary values. In our computations we have not been able to push the numerics too much in that direction, but as the number of zeros increases, we do see that the match improves.

We will conclude by taking stock of what we expect from the Hamiltonian for AdS$_2$ quantum gravity, what we know from the spectrum of SYK and SYK-like tensor models, and what is  known/expected about the Riemann Hamiltonian. Let us first see what we expect from a holographic dual of AdS$_2$:

\begin{itemize}
\item[0.] It should be a quantum mechanics theory (0+1 dimenions)

\item[1.] We want to have a well-defined large-$N$ limit, so that there is some notion of semi-classical spacetime limit.

\item[2.] We want the entropy to scale as $N$ to some positive power, so that we get finite entropy at zero temperature. This loosely means that there should be exponentially large in $N$ degeneracy in the ground state (or at least there should be enough density of states near the ground state, which is what happens in SYK and related models). This expectation is based on the fact that we would like AdS$_2$ to be a model for near horizon geometry of extremal black holes, which have non-zero entropy\footnote{It is not very clear to us why the same logic does not apply to higher dimensional extremal black branes, where the corresponding quantum gravity theories do not have huge degeneracies around ground state. The difference is possibly related to the fact that AdS$_2$ can be understood as a ``very" near horizon limit.}.

\item[3.] There should be chaos, and it should saturate the chaos bound.

\item[4.] There is an emergent conformal symmetry in the infrared (which could be spontaneously, and weakly explicitly, broken).

\end{itemize}
The SYK and related models are believed to satisfy these features. But together with that, they satisfy a few extra features:

\begin{itemize}
\item[3a.] The spectrum exhibits features of random matrices.
\item[5.] The ground state energy scales linearly with $N$. 
\item[6.] Their spectrum satisfies a spectral mirror symmetry, after disorder averaging in the disorder averaged models, and without any disorder average in many of the tensor models \cite{KSS, KKS} and, as we have seen, in one instance (at least) of the generalized SYK model. 
\end{itemize}
Now in the case of Riemann zeros, the situation is as follows:
\begin{itemize}
\item[0.] Yes, it is a 0+1 theory.

\item[1.] The zero-number provides a natural large-$N$ limit (note that the number of the zero need not directly be the $N$, but could be some function of it.)

\item[2.] The density of states is flat, there is no approximate Wigner-like distribution near the edges if we truncate at some finite zero. But note however from our plot of the Riemann SFF, that the plateau height is (roughly) exponentially smaller than 1. This could be a suggestion that there is entropy in the Riemann zeros, in an appropriate simultaneous zero temperature, large-$N$ limit. Note that in the results of \cite{Cotler} also, to get finite zero temperature  entropy (more precisely $S/N$) using finite-$N$ data, one had to extrapolate to the large-$N$ limit.


\item[3/3a.] There is certainly chaos, because of the connection to RMT and Gutzwiller trace formula, level repulsion, pair correlation, etc \cite{Keating}. Checking for maximal chaos, is usually done via computing four point Out of Time Ordering Correlators (OTOC). In the case of Riemann, we only know the eigenvalues, not the eigenvectors. This raises the interesting question: If we just know eigenvalues and not the eigenvectors (if we knew both, we could reconstruct the Hamiltonian trivially), can we evaluate the Lyapunov exponent (or OTOC) of the system? Some interesting developments, on the relationship between the SFF and the OTOC in RMTs, has been recently discussed in \cite{Cotler:2017jue}. 

\item[4.] Again, because we only know the eigenvalue spectrum, it is hard to say if there is emergent conformal symmetry in the infrared. It will again be interesting to investigate the conditions under which a given eigenvalue spectrum leads to IR conformal invariance. But note the fact that the SFF for SYK, before the dip time, goes like $\frac{1}{t^3}$ is directly related to the fact that the low energy excitations are described by the Schwarzian action, as pointed out in \cite{Cotler}. It has been observed, in \cite{Maldacena:2016upp} and \cite{Engelsoy:2016xyb}, that the Schwarzian action has a clear counterpart in AdS$_2$ dilaton gravity. Hence, if a gravity dual of the Rieman zeros exists, it is unlikely to be described by AdS$_2$ dilaton gravity on the dual side.

\item[5.] The $w_n$'s do scale approximately linearly with $n$.

\item[6.] The non-trivial zeros of Riemann are symmetric around the real axis, which guarantees that the Riemann Hamiltonian has spectral mirror symmetry. 
\end{itemize}

We will conclude with one final comment. The density of states in the  ungauged tensor models are distinct in qualitative ways from that in the disorder averaged SYK models \cite{KSS, KKS}. In particular, they exhibit huge degeneracies in the spectrum, which must clearly be related to the fact that the theory has a huge symmetry. It will be interesting to see how gauging the theory will affect the spectrum: in particular, it will be very interesting to compare the spectra of the various theories (including Riemann zeros) after the tensor models are gauged. 
Some work towards solving for the spectrum of the gauged theory is currently under way.


\section{Hints of level repulsion in the GW model?}

We want to make some comments on the first hints of level repulsion in the GW model for the particular case of $n=2$,$D=3$ we are studying. In \cite{KSS}, the authors applied the standard diagnostic given by the unfolded level spacing statistic and found clear evidence of level repulsion. But the numerosity was too low to convincingly fit the Wigner surmise, but it is believable that the level repulsion will become more and more evident by increasing the numerosity. 

In this Section we instead study the same question using another kind of diagnostic: the r-statistics. We will see that the results of the r-statistics do {\it not} agree with the results of unfolded level spacing statistics, and we will try to explain the reason for this discrepancy. 

\begin{figure}
\centering
    \includegraphics[width=0.6\textwidth]{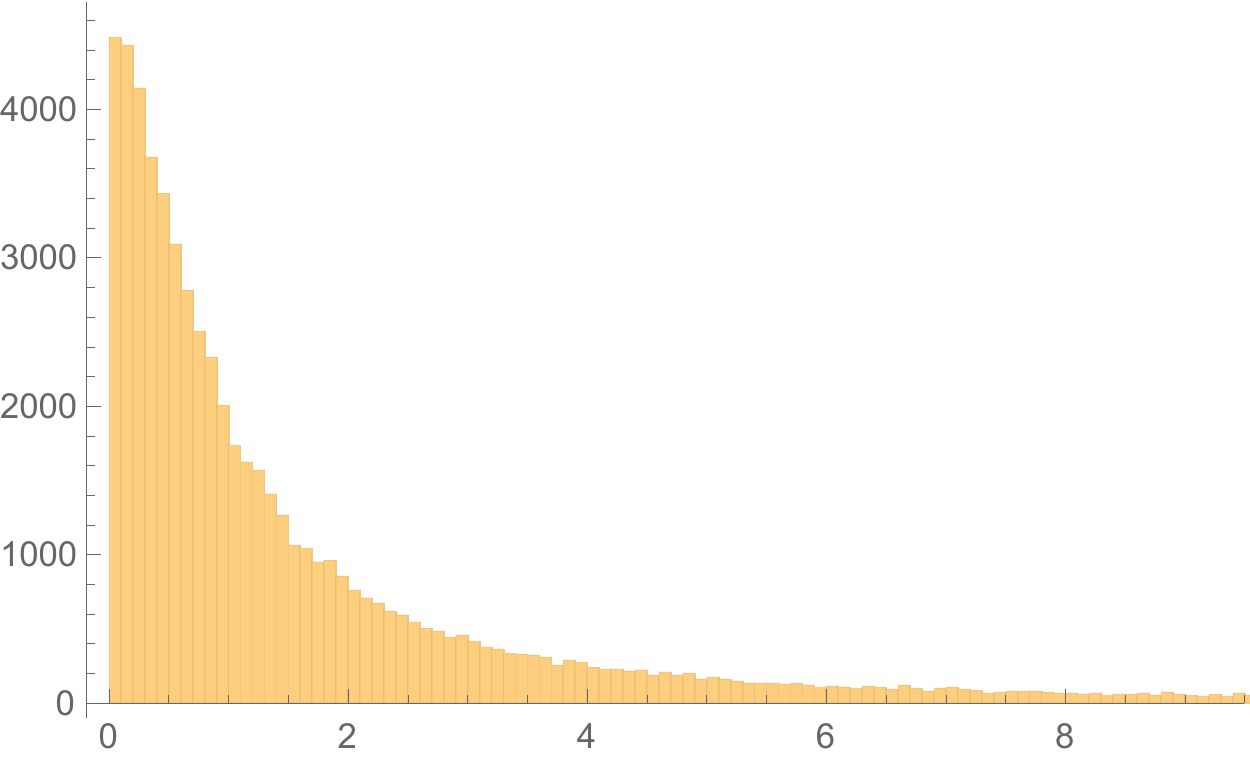}
   \caption{The ratio of the distances between nearest eigenvalues for the SYK spectrum given in Figure \ref{fig:SYK_spectrum_full}.}
    \label{fig:rSYK_naive}
\end{figure}

We will use the recipe explained in \cite{You:2016ldz}. We construct the list of values of the energy eigenstates $\{E_n \}$ and we order them in ascending order $E_1 < E_2 < \cdots$. Then we construct the level spacings between adjacent non degenerate eigenvalues, $\Delta \, E_n = E_n - E_{n+1}$. This quantities are not appropriate to study the presence of repulsion between adjacent eigenvalues, since the dependence on the average densisty of states can affect the estimates. Hence, we introduce the ratios $r_n = \Delta \,E_n / \Delta \, E_{n+1}$ which remove the dependence on the average density. For an integrable system, the distribution of these ratios tends to the following distribution
\begin{align}
\label{eq:poissondistribution}
P(r) = \frac{1}{(1+r)^2} ,
\end{align}
whereas for RMT the distribution approaches the Wigner surmise
\begin{align}
\label{eq:wignersurmise}
P (r) = \alpha \frac{(r + r^2)^\beta}{(1 + r + r^2)^{1 + 3 \beta / 2}}
\end{align}
where $\alpha$ is a numerical constant and $\beta = 1, \, 2, \, 4$ for GOE, GUE and GSE, respectively. As a warm-up, we have computed the distribution $P(r)$ for the SYK spectrum. The result is plotted in Figure \ref{fig:rSYK_naive}.
Contrary to the expectation $P(r)$ is following the curve (\ref{eq:poissondistribution}) which is typical for an integrable system. The reason  is that the SYK spectrum is given by states with different fermion parities and the SYK hamiltonian does preserve the parity symmetry. Hence, states with different fermion parities are not correlated, they do not repel each other and the r-statistics does not show repulsion.

\begin{figure}
\centering
    \includegraphics[width=0.6\textwidth]{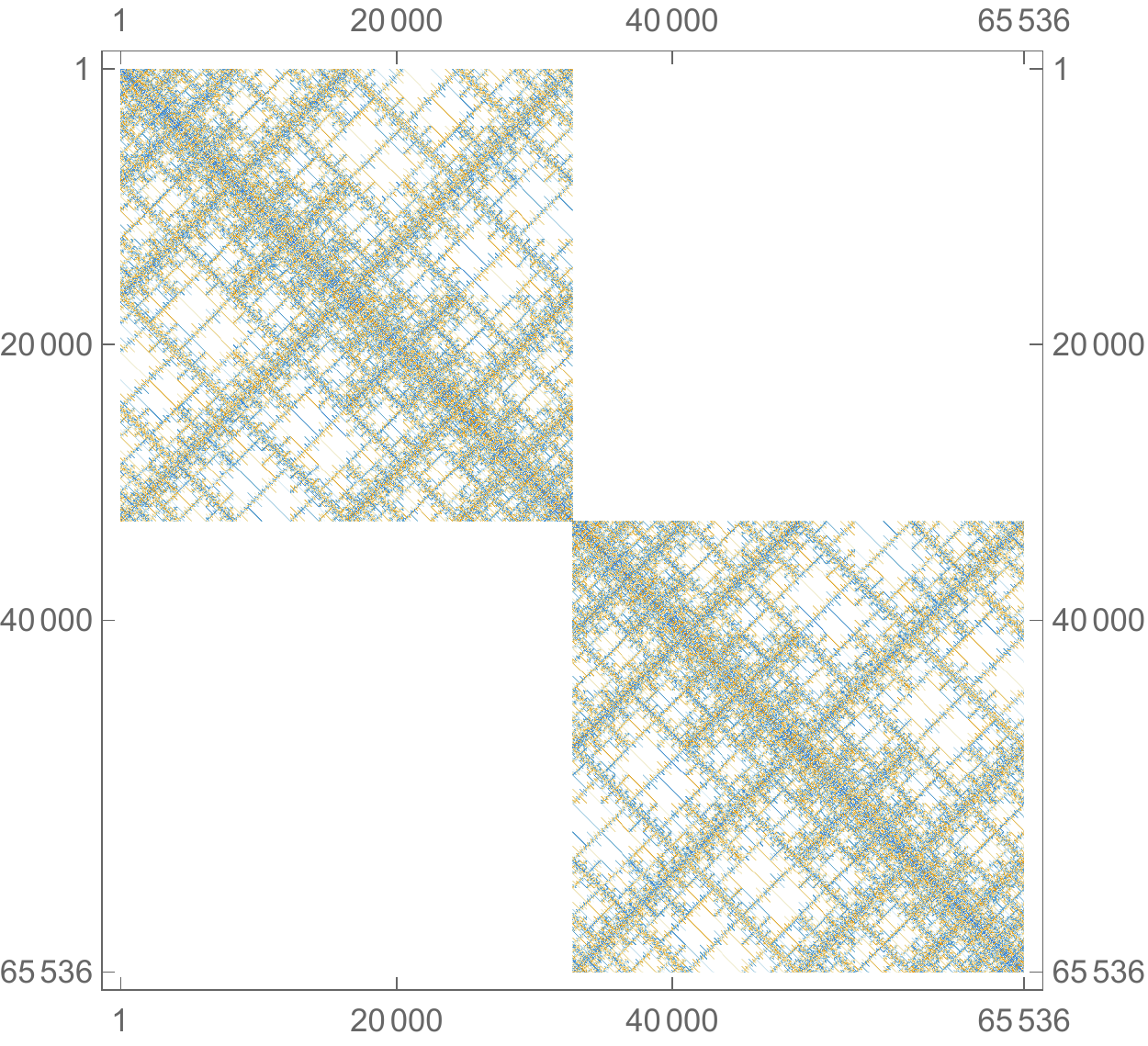}
   \caption{The hamiltonian of the SYK model in the block diagonal form.}
    \label{fig:SYK_improved}
\end{figure}

\begin{figure}
\centering
    \includegraphics[width=0.6\textwidth]{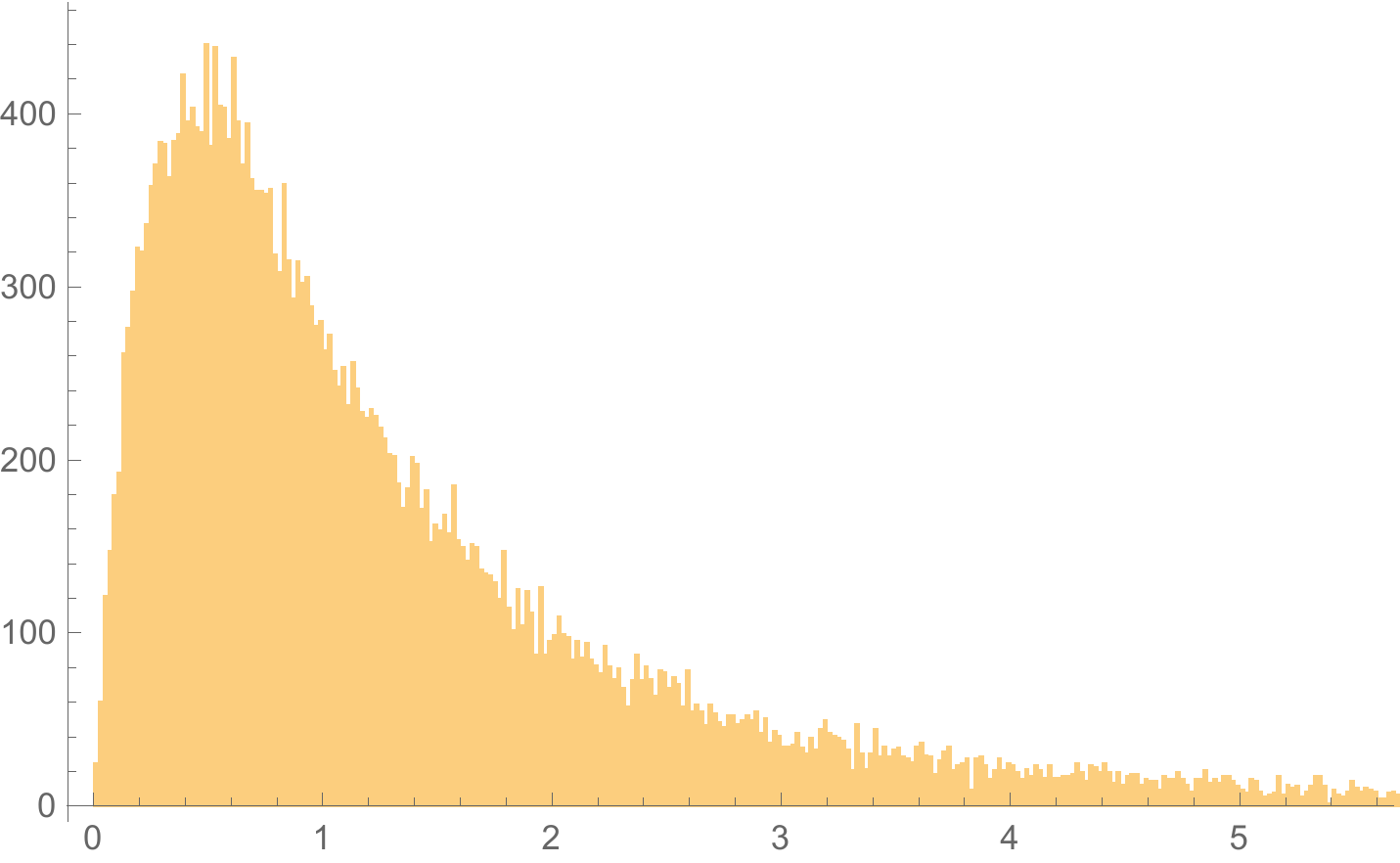}
   \caption{The ratio of the distances between nearest eigenvalues for the upper block of the SYK hamiltonian.}
    \label{fig:rSYK_improved_upper}
\end{figure}

To solve this problem, we perform the spectral analysis among the states with only positive (or negative) chiralities. To separate these two subsystems, we choose the following set of Gamma matrices to represent the fermions  \cite{Garcia-Garcia:2016mno}. We start with the definitions
\begin{align}\label{eq:gammaimprovedone}
\gamma_1^{(2)} = \sigma_1 \ , \qquad \gamma_2^{(2)} = \sigma_2 \ , \qquad \gamma_3^{(2)} = \sigma_3 \ ,
\end{align}
and then we define the Gamma mattrices in $N= d + 2$ dimensions using the recursion relations
\begin{align}\label{eq:gammaimprovedtwo}
& \gamma_k^{(d+2)} = \sigma_1 \otimes \gamma_k^{(d)} \ , \qquad k= 1 \, , \cdots \, , d+1 \ , \nonumber \\
& \gamma_{d+2}^{(d+2)} = \sigma_2 \otimes 1_{2^{d/2}} \ .
\end{align}
The SYK hamiltonian, when expressed in terms of the gamma matrices just introduced, takes the block diagonal form depicted in Figure  \ref{fig:SYK_improved}. 
The two blocks are  eigenstates of the chiral operator. Hence we analyse both separately. We have evaluated the probabilities $P(r)$ for both the upper block hamiltonian and the lower block hamiltonian and the resulting plot for the upper block is presented in Figure \ref{fig:rSYK_improved_upper} (the plot for the lower block is very similar). We see clearly the repulsive behavior at very low distances. The two plots respects the Wigner surmise behavior as expexted.

\begin{figure}
\centering
    \includegraphics[width=0.6\textwidth]{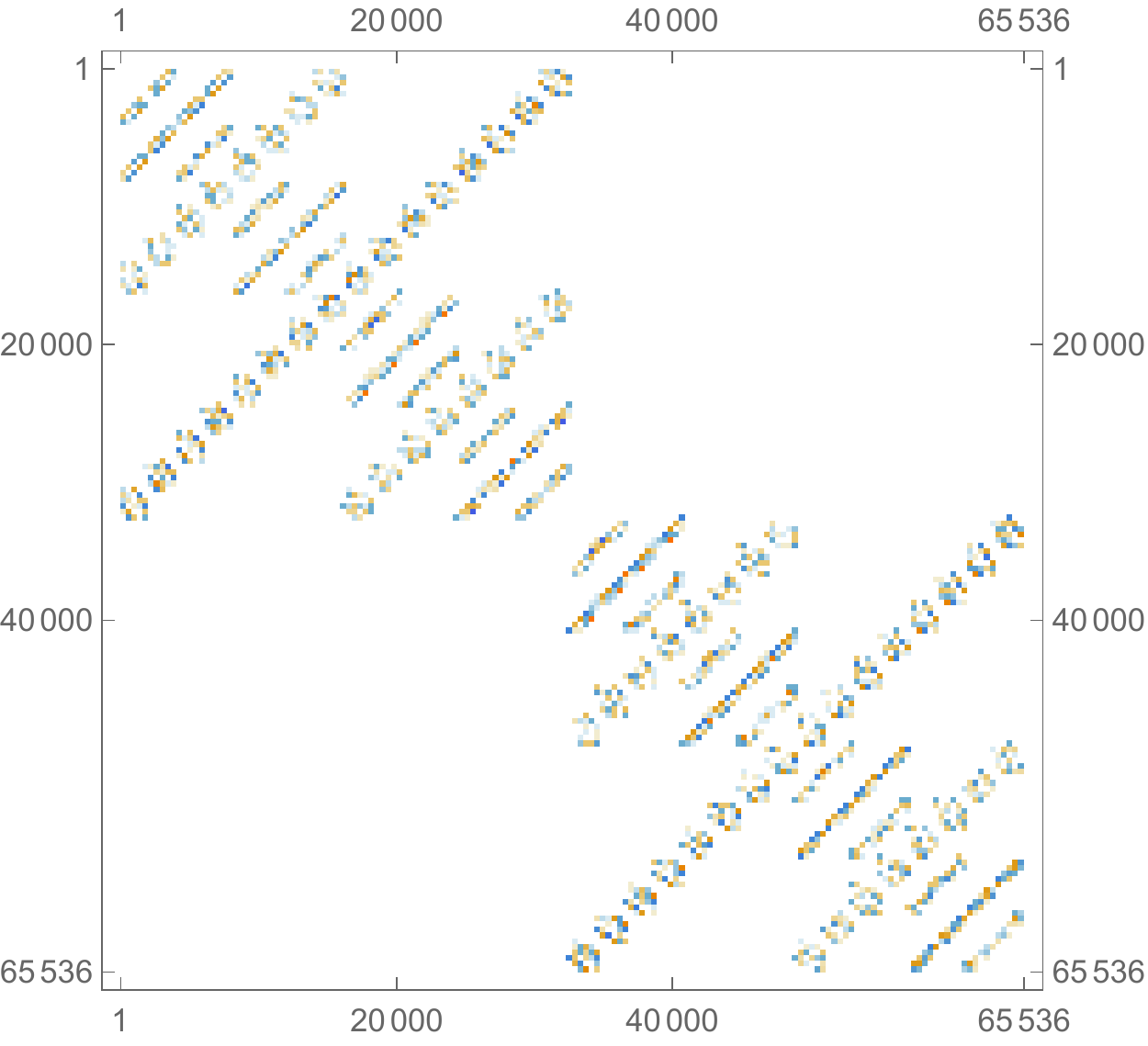}
   \caption{The hamiltonian of the GW model in the block diagonal form.}
    \label{fig:GW_improved}
\end{figure}

\begin{figure}
\centering
    \includegraphics[width=0.6\textwidth]{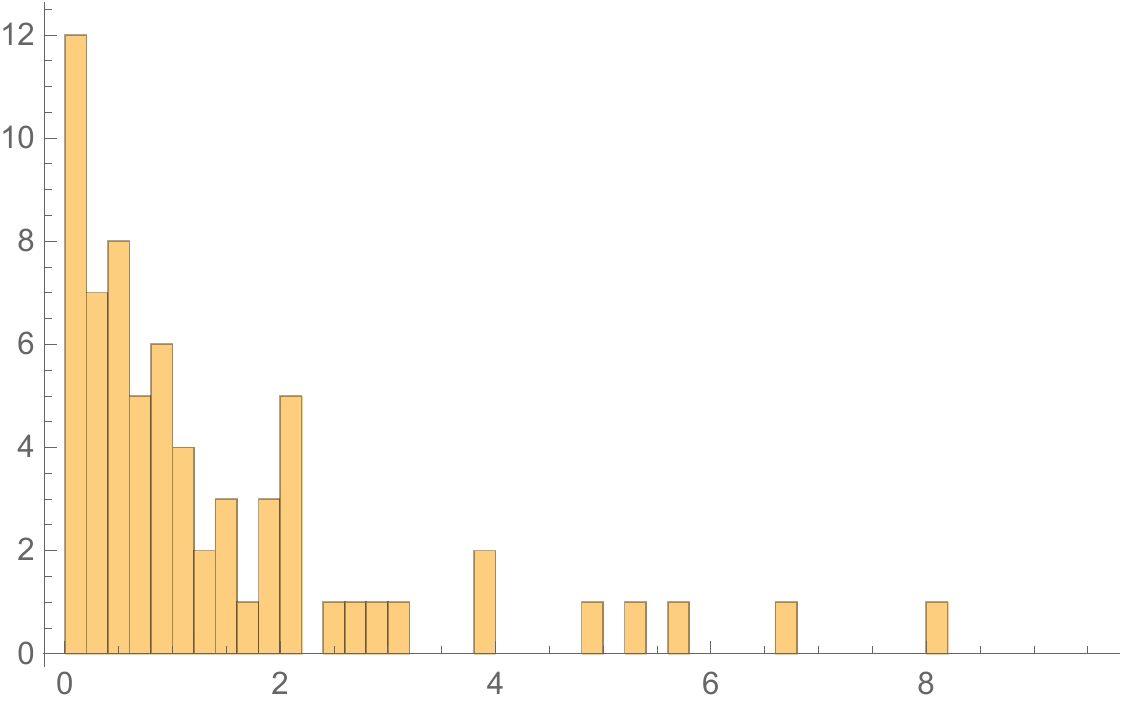}
   \caption{The ratio of the distances between nearest eigenvalues for the upper block of the GW hamiltonian.}
    \label{fig:rGW_improved_upper}
\end{figure}

We now move to consider the GW model. Once again, in terms of the Gamma matrices (\ref{eq:gammaimprovedtwo}), the hamiltonian takes a block diagonal form which we present in Figure \ref{fig:GW_improved}.
 Our interest is again in computing the probabilities $P(r)$ for the upper block (once again, the results for the lower block are very similar). We have computed them and they are presented in Figure \ref{fig:rGW_improved_upper}.
This plot does not show any hints of repulsion: in other words it does not agree with the results of the unfolded level spacing statistics.

\begin{figure}
\centering
    \includegraphics[width=0.6\textwidth]{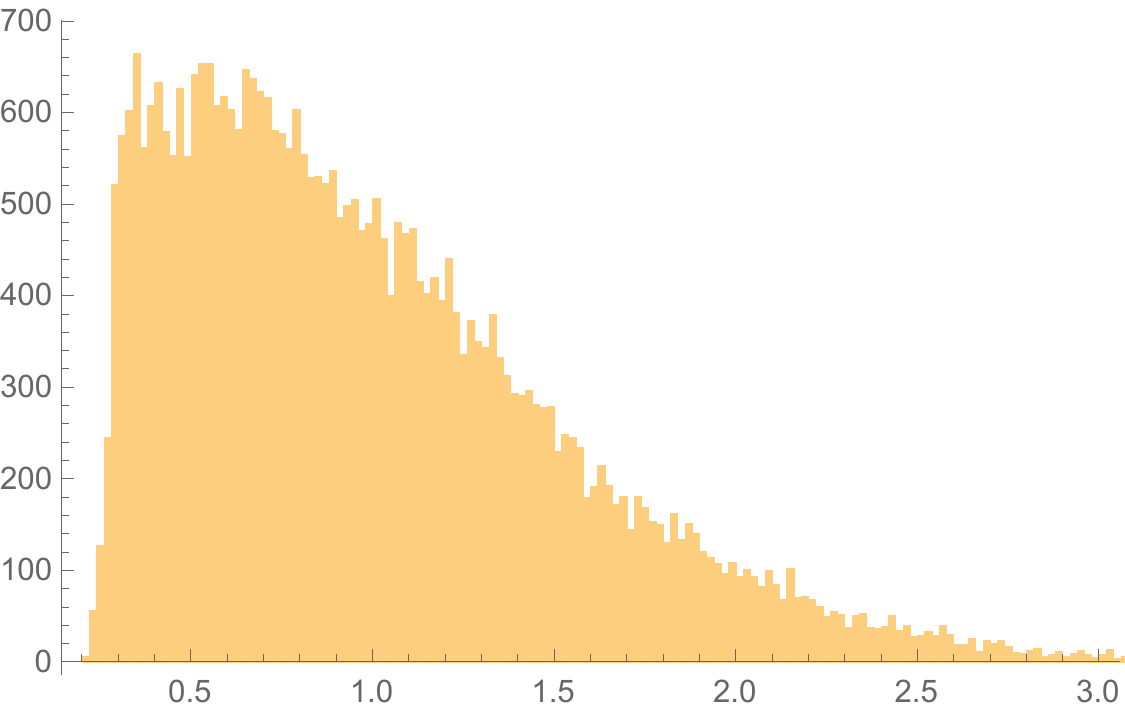}
   \caption{The distribution of the unfolded level spacing for the full SYK Hamiltonian.}
    \label{fig:unfolded_level_SYK_naive}
\end{figure}

Let us try to understand the reason of such a discrepancy between the results of the unfolded level spacing and the results of the r-statistics. To guide our intuition, we plot in Figure  \ref{fig:unfolded_level_SYK_naive} the unfolded level spacing distribution for the {\it full} SYK Hamiltonian, i.e. without separating the two blocks with different chiralities.
The important feature of this figure is that, even if the spectrum is not separated by the different symmetry sector, the hints of level repulsion are evident. By comparing Figure \ref{fig:unfolded_level_SYK_naive} with Figure \ref{fig:rSYK_naive} we conclude that the unfolded level spacing is {\it less} sensitive than the r-statistics to the presence of different symmetry sectors in the Hamiltonian. In other words, the former is a {\it more} robust predictor of level repulsion when there are symmetry sectors in the theory.

Given this observation we can go back to the case of the GW model: in this model there is an additional $O(n)^6$ global symmetry. Hence, it is reasonable to conjecture that to see the level repulsion with r-statistics, one should separate the spectrum according to the different chiralities {\it and} to the different $O(n)^6$ sectors. In particular, one should work with the different multiplets of $O(n)^6$ into which the states in the Hilbert space break up, separately. Unfortunately, if one tries to follow this route, the numerosity become too small, so to competely settle this issue one will have to go to higher $n$.

Summarizing, we believe that the tension, for the GW model, between the results of the unfolded level spacing and the r-statistics is due to the fact that the r-statistics is more sensitive to the presence of different symmetry sectors in the GW Hamiltonian, and that the results obtained with the unfolded level spacings in \cite{KSS} are really first hints of level repulsion.

\providecommand{\href}[2]{#2}

\end{document}